\documentclass[graybox,envcountchap,sectrefs]{svmult}

\usepackage{mathptmx}
\usepackage{helvet}
\usepackage{courier}
\usepackage{braket} 
\usepackage{algorithmicx}
\usepackage{algpseudocode} 
\usepackage{amsfonts}

\usepackage{type1cm}         
\usepackage{exercise}
\usepackage{makeidx}         
\usepackage{graphicx}        
\usepackage{multicol}        
\usepackage[bottom]{footmisc}
\usepackage[usenames,dvipsnames,x11names]{xcolor}
 \usepackage{listings}
 \usepackage{epic}
 \usepackage{eepic}
 \usepackage{a4wide}
 \usepackage{color}
 \usepackage{amsmath}
 \usepackage{amssymb}
 \usepackage[T1]{fontenc}
 \usepackage{cite} 
 \usepackage{shadow}
 \usepackage{hyperref}
 \usepackage{bezier}
 \usepackage{pstricks}
\usepackage{textcomp,type1ec,pdfpages}
\usepackage{bera}

\makeindex             

\begin{document}

\newcommand {\bea}{\begin{eqnarray}}
\newcommand {\eea}{\end{eqnarray}}
\newcommand {\be}{\begin{equation}}
\newcommand {\ee}{\end{equation}}
\newcommand {\qslash}{q\!\!\!/}
\newcommand {\muslash}{\mu\!\!\!/}
\newcommand {\Dslash}{D\!\!\!/}
\newcommand {\bc}{\begin{center}}
\newcommand {\ec}{\end{center}}
\def\lsim{\mathrel{\rlap{\lower4pt\hbox{\hskip1pt$\sim$}}
    \raise1pt\hbox{$<$}}}               
\def\gsim{\mathrel{\rlap{\lower4pt\hbox{\hskip1pt$\sim$}}
    \raise1pt\hbox{$>$}}}

\title{Quantum Chromodynamics}
\author{Thomas Sch\"afer}
\institute{Thomas Sch\"afer \at Department of Physics, 
North Carolina State University, Raleigh, NC 27695,  USA, 
\email{tmschaef@ncsu.edu}}

\maketitle
\abstract{We present a brief introduction to QCD, the QCD phase
diagram, and non-equilibrium phenomena in QCD. We emphasize
aspects of the theory that can be addressed using computational
methods, in particular euclidean path integral Monte Carlo, 
fluid dynamics, kinetic theory, classical field theory and 
holographic duality.}

\section{Introduction}

 The goal of this chapter is to provide a brief summary of Quantum 
Chromodynamics (QCD) and the QCD phase diagram, and to give an introduction 
to computational methods that are being used to study different aspects of 
QCD. Quantum Chromodynamics is a remarkable theory in many respects. QCD 
is an almost parameter free theory. Indeed, in the context of nuclear 
physics QCD is completely characterized by the masses of the up, down,
and strange quark, and a reasonable caricature of nuclear physics emerges
in the even simpler case in which the up and down quark are taken to
be massless, and the strange quark is infinitely heavy. QCD nevertheless
accounts for the incredible richness of the phase diagram of strongly 
interacting matter. QCD describes finite nuclei, normal and superfluid 
states of nuclear matter, color superconductors, hadronic gases, quark 
gluon plasma, and many other states. This rich variety of states is
reflected in the large number of computational methods that have been 
brought to bear on problems in QCD. This includes a large number 
of methods for the structure and excitations of finite Fermi systems, 
quantum Monte Carlo methods, and a variety of tools for equilibrium
and non-equilibrium statistical mechanics. 

 The bulk of this book is devoted to the study of few and many nucleon
systems. Summarizing everything else in one brief chapter is obviously
out of the question, both because of limitations of space and because
of my limited expertise. I will therefore be very selective, and focus 
on a number of very simple yet powerful ideas. This reflects, in part, 
my background, which is not primarily in computational physics. It 
also reflects my conviction that progress in computational physics is 
unfortunately often reflected in increasingly complicated codes that 
obscure the simplicity of the underlying methods. 

\section{Path integrals and the Metropolis algorithm}
\label{sec_qm}

 Consider a simple quantum mechanical problem, the motion of a particle
in a one-dimensional potential. In order to be specific I will focus
on the double well potential $V(x)=\lambda(x^2-\eta^2)^2$, where $\eta$ 
and $\lambda$ are parameters. The Hamiltonian is
\be 
\label{H_dw}
 H = \frac{p^2}{2m}+\lambda (x^2-\eta^2)^2\, . 
\ee
Using a change of variables I can set $2m=\lambda=1$. This implies that
there is only one physical parameter in this problem, the barrier 
separation $\eta$. The regime $\eta\gg 1$ corresponds to the limit 
in which the system has two almost degenerate minima that are split
by semi-classical tunneling events. The energy eigenstates and wave 
functions are solutions of the eigenvalue problem $H|n\rangle = |n\rangle 
E_n$. Once the eigenstates are known I can compute all possible 
correlation functions 
\be 
\Pi_n(t_1,t_2,\ldots,t_n) = \langle 0 |x(t_1) x(t_2) \ldots
 x(t_n)|0\rangle \, ,
\ee
by inserting complete sets of states. An alternative to the Hamiltonian 
formulation of the problem is the Feynman path integral \cite{Feynman}. 
The path integral for the anharmonic oscillator is given by 
\be
\label{pathint}
 \langle x_1| e^{-iHt_f}|x_0\rangle = 
  \int_{x(0)=x_0}^{x(t_f)=x_1} {\cal D}x\, e^{iS}, 
  \hspace{1cm}
  S=\int_0^{t_f} dt\, \left(
   \frac{1}{4}\dot{x}^4-(x^2-\eta^2)^2 \right).
\ee
This expression contains a rapidly oscillating phase factor $e^{iS}$,
which prohibits any direct numerical attempt at computing the path 
integral. The standard approach is based on analytic continuation
to imaginary time $\tau=it$. This is also referred to as Euclidean
time, because the Minkowski interval $dx^2-dt^2$ turns into the 
Euclidean expression $dx^2+d\tau^2$. In the following I will consider 
the euclidean partition function
\be
\label{z}
 Z(T) = \int {\cal D}x\, e^{-S_E}, \hspace{1cm}
  S_E=\int_0^{\beta} d\tau\, \left(
   \frac{1}{4}\dot{x}^4+(x^2-\eta^2)^2 \right),
\ee
where $\beta=1/T$ is the inverse temperature and we assume periodic 
boundary conditions $x(0)=x(\beta)$. To see that equ.~(\ref{z})
is indeed the partition function we can use equ.~(\ref{pathint}) to 
express the path integral in terms of the eigenvalues of the Hamiltonian, 
$Z(T)=\sum_n\exp(-E_n/T)$. 
In the following I will describe numerical simulations using a discretized 
version of the euclidean action. For this purpose I discretize the euclidean 
time coordinate $\tau_j=ja,\,i=1,\ldots n$ where $a=\beta/n$ is the length 
of time interval. The discretized action is given by
\be
\label{S_disc}
 S = \sum_{i=1}^{n}\left\{ \frac{1}{4a} (x_i-x_{i-1})^2
  + a(x_i^2-\eta^2)^2 \right\},
\ee
where $x_i=x(\tau_i)$. I consider periodic boundary conditions $x_0=x_n$. 
The discretized euclidean path integral is formally equivalent to the 
partition function of a statistical system of (continuous) ``spins'' 
$x_i$ arranged on a one-dimensional lattice. This statistical system 
can be studied using standard Monte-Carlo sampling methods. In the following 
I will use the Metropolis algorithm \cite{Metropolis:1953am}. Detailed
numerical studies of the euclidean path integral can be found in 
\cite{Creutz:1980gp,Shuryak:1984xr,Shuryak:1987tr,Schafer:2004xa}.

 The Metropolis method generates an ensemble of configurations
$\{x_i\}^{(k)}$ where $i=1,\ldots, n$ labels the lattice points and $k=
1,\ldots,N_{conf}$ labels the configurations. Quantum mechanical averages 
are computed by averaging observables over many configurations, 
\be 
\langle {\cal O} \rangle = \lim_{N_{conf}\to\infty}
 \frac{1}{N_{conf}}\sum_{k=1}^{N_{conf}}
 {\cal O}^{(k)}
\ee
where ${\cal O}^{(k)}$ is the value of the classical observable
${\cal O}$ in the configuration $\{x_i\}^{(k)}$. The configurations 
are generated using Metropolis updates $\{x_i\}^{(k)}\to \{x_i\}^{(k+1)}$. 
The update consists of a sweep through the lattice during which a trial 
update $x_i^{(k+1)}= x_i^{(k)} +\delta x$ is performed for every lattice 
site. Here, $\delta x$ is a random number. The trial update is accepted 
with probability
\be
 P\left(x_i^{(k)}\to x_i^{(k+1)}\right)=
   \min\left\{\exp(-\Delta S),1\right\},
\ee 
where $\Delta S$ is the change in the action equ.~(\ref{S_disc}). 
This ensures that the configurations $\{x_i\}^{(k)}$ are distributed 
according the ``Boltzmann'' distribution $\exp(-S)$. The distribution 
of $\delta x$ is arbitrary as long as the trial update is micro-reversible, 
i.~e.~is equally likely to change $x_i^{(k)}$ to $x_i^{(k+1)}$ and back. 
The initial configuration is arbitrary. In order to study equilibration 
it is useful to compare an ordered (cold) start with $\{x_i\}^{(0)}=
\{\eta\}$ to a disordered (hot) start $\{x_i\}^{(0)}=\{r_i\}$, where 
$r_i$ is a random variable. 

 The advantage of the Metropolis algorithm is its simplicity and robustness. 
The only parameter to adjust is the distribution of $\delta x$. A simple
choice is to take $\delta x$ to be a Gaussian random number, and choose the
width of the distribution so that the average acceptance rate for the trial 
updates is around $50\%$. Fluctuations of ${\cal O}$ provide an estimate 
in the error of $\langle {\cal O}\rangle$. The uncertainty is given by 
\be 
\Delta \langle {\cal O} \rangle =
 \sqrt{\frac{\langle {\cal O}^2\rangle -\langle{\cal O}\rangle^2}
            {N_{conf}}}. 
\ee
This requires some care, because the error estimate is based on 
the assumption that the configurations are statistically independent. 
In practice this can be monitored by computing the auto-correlation 
``time'' in successive measurements ${\cal O}(\{x_i\}^{(k)})$. 

 I have written a simple fortran code that implements the Metropolis 
algorithm for euclidean path integrals \cite{Schafer:2004xa}. The 
most important part of that code is a sweep through the lattice 
with a Metropolis update on every site $\tau_j$: 

\vspace*{0.3cm} 
\begin{lstlisting}
 do j=1,n-1
            
         nhit = nhit+1  
          
         xpm = (x(j)-x(j-1))/a
         xpp = (x(j+1)-x(j))/a
         t = 1.0/4.0*(xpm**2+xpp**2)
         v = (x(j)**2-f**2)**2
         sold = a*(t+v)

         xnew = x(j) + delx*(2.0*ran2(iseed)-1.0)
             
         xpm = (xnew-x(j-1))/a
         xpp = (x(j+1)-xnew)/a
         t = 1.0/4.0*(xpm**2+xpp**2)
         v = (xnew**2-f**2)**2
         snew = a*(t+v)
         dels = snew-sold           
                       
         p  = ran2(iseed)          
         if (exp(-dels) .gt. p) then
            x(j) = xnew
            nacc = nacc + 1
         endif
            
enddo
\end{lstlisting}

\begin{figure}[t]
\begin{center}
\includegraphics[width=9cm]{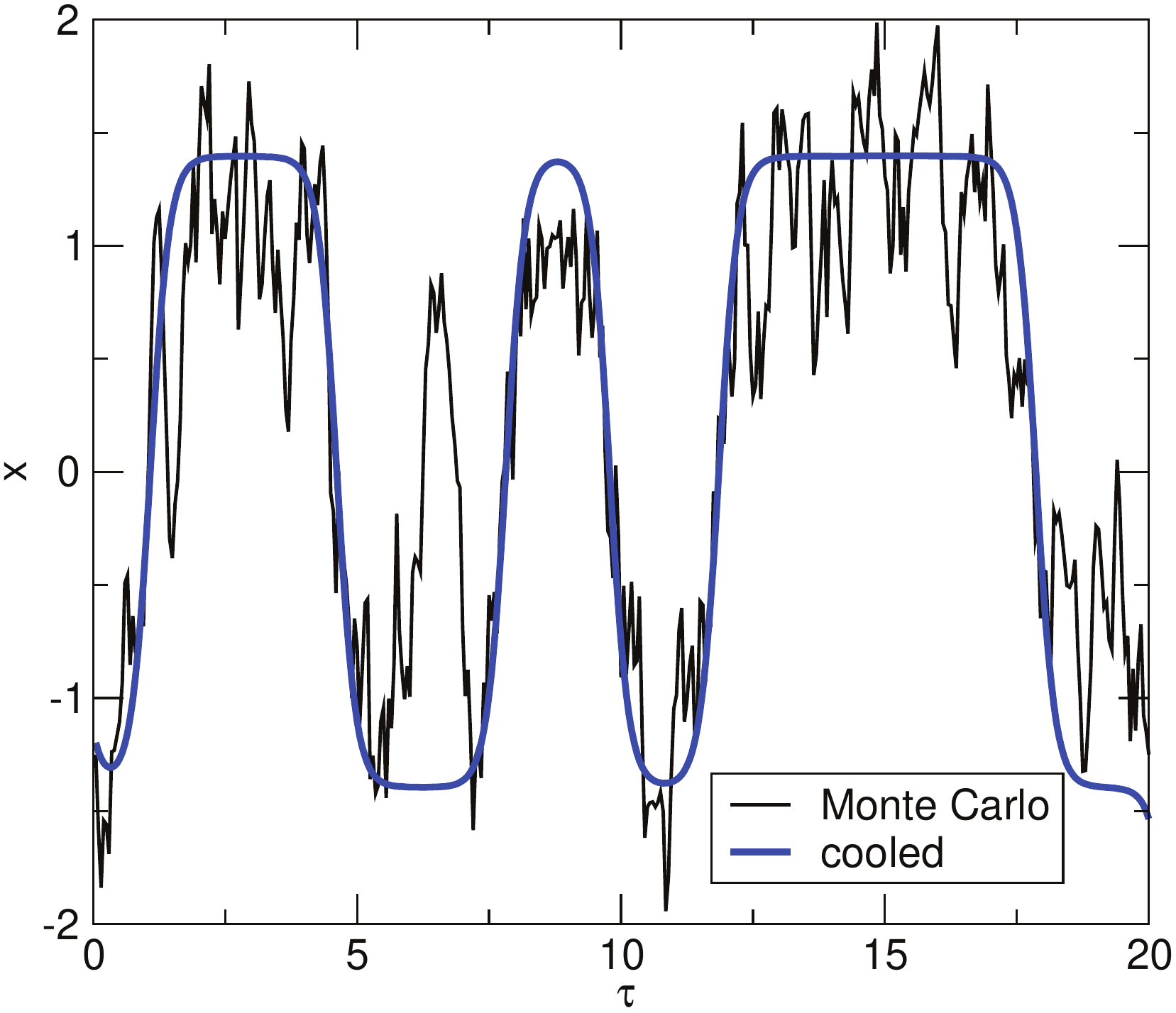}  
\end{center}  
\caption{\label{fig_path}
Typical euclidean path obtained in a Monte Carlo simulation
of the discretized euclidean action of the double well 
potential for $\eta=1.4$. The lattice spacing in the 
euclidean time direction is $a=0.05$ and the total number 
of lattice points is $N_\tau=800$. The green curve shows 
the corresponding smooth path obtained by running 100 
cooling sweeps on the original path. }
\end{figure}

\vspace*{0.3cm} 
 Here, {\tt sold} is the local action corresponding to the initial
value of ${\tt x(j)}$, and {\tt snew} is the action after the trial
update. The trial update is accepted if {\tt exp(-dels)} is greater
that the random variable ${\tt p}$. The function {\tt ran2()} generates 
a random number between 0 and 1, and {\tt nacc/nhit} measures the 
acceptance rate. A typical path is shown in Fig.~\ref{fig_path}. An 
important feature of the paths in the double well potential is the 
presence of tunneling events. Indeed, in the semi-classical regime 
$\eta\gg 1$, a typical path can be understood as Gaussian fluctuations 
superimposed on a series of tunneling events (instantons). 

 The path integral method does not provide direct access to the eigenvalues
of the Hamiltonian, but it can be used to compute imaginary time correlation
functions
\be
\label{qm_cor}
 \Pi_n^E(\tau_1,\ldots,\tau_n)=\langle x(\tau_1)\ldots x(\tau_n)\rangle.
\ee
Note that the average is carried out with respect to the partition 
function in equ.~(\ref{z}). In the limit $\beta\to\infty$ this 
corresponds to the ground state expectation value. A very important 
observable is the two-point function $\Pi^E(\tau)\equiv \Pi^E_2(0,\tau)$. 
The euclidean correlation functions is related to the eigenstates of 
the Hamiltonian via a spectral representations. This representation 
is obtained by inserting a complete set of states into 
equ.~(\ref{qm_cor}). The result is
\be 
\label{spec}
\Pi^E(\tau)= \sum_n |\langle 0|x|n\rangle|^2 
 \exp(-(E_n-E_0)\tau),
\ee
where $E_n$ is the energy of the state $|n\rangle$. This can be written as
\be
\Pi^E(\tau)= \int dE\, \rho(E) \exp(-(E-E_0)\tau),
\ee
where $\rho(E)$ is the spectral function. In the case of the double well
potential there are only bound states and the spectral function is a sum 
of delta-functions. Equ.~(\ref{spec}) shows that the euclidean correlation 
function is easy to construct once the energy eigenvalues and eigenfunctions 
are known. The inverse problem is well defined in principle, but numerically 
much more difficult. The excitation energy of the first excited state 
$\Delta E_1 = E_1-E_0$ is easy to extract from the exponential decay of the 
two-point functions, but higher states are more difficult to compute. A 
technique for determining the spectral function from 
euclidean correlation functions is the maximum entropy image reconstruction 
method, see \cite{Jarrell:1996rrw,Asakawa:2000tr}.

 The calculation of correlation functions in a Monte Carlo simulation
is very straightforward. All I need to do is multiply the values of 
$x(\tau_i)$ for a given path, and then average over all paths:
 
\vspace*{0.3cm} 
\begin{lstlisting}
do ic=1,nc
            
      ncor = ncor + 1 
      ip0 = int( (n-np)*ran2(iseed) ) 
      x0  = x(ip0) 
            
      do ip=1,np
         x1 = x(ip0+ip)
         xcor = x0*x1
         x2cor= xcor**2
         xcor_sum(ip)  = xcor_sum(ip)  + xcor
         xcor2_sum(ip) = xcor2_sum(ip) + xcor**2
   enddo  
enddo
\end{lstlisting}

\vspace*{0.3cm} 
 The advantages of this method are that it is extremely robust, that it 
requires no knowledge (or preconceived notion) of what the wave function 
looks like, and that it can explore a very complicated configuration space.
On the other hand, in the case of one-dimensional quantum mechanics, the 
Metropolis method is very inefficient. Using direct diagonalization in a 
finite basis it is not difficult to compute the energies of the first 
several states in the potential in equ.~(\ref{H_dw}) with very high accuracy, 
$\Delta E/E_0 \sim O(10^{-6})$ or better. On the other hand, using the 
Monte Carlo method, it is quite difficult to achieve an accuracy of 
$O(10^{-2})$ for observable other than $(E_1-E_0)/E_0$. The advantage of the 
Monte Carlo method is that the computational cost scales much more 
favorably in high dimensional systems, such as quantum mechanics of 
many particles, or quantum field theory.

 The Monte Carlo method also does not directly provide the ground state
energy, or the partition function and free energy at finite temperature.
In quantum mechanics we can compute the ground state energy from the 
expectation value of the Hamiltonian $\langle H\rangle = \langle T+V
\rangle$ in the limit $\beta\to\infty$. The expectation value of the kinetic 
energy is singular as $a\to 0$, but this problem can be overcome by using
the Virial theorem
\be 
\langle H\rangle = \left\langle \frac{x}{2}V'+V \right\rangle\, . 
\ee
There is no simple analog of this method in quantum field theory. A method 
for computing the free energy which does generalize to quantum field theory
is the adiabatic switching technique. The idea is to start from a reference 
system for which the free energy is known and calculate the free energy 
difference to the real system using Monte Carlo methods. For this purpose 
I write the action as 
\be 
S_\alpha=S_0+\alpha\Delta S\, , 
\ee
where $S_0$ is the action of the reference system, $\Delta S$ is defined 
by $\Delta S=S-S_0$ where $S$ is the full action, and $\alpha$ can be viewed 
as a coupling constant. The action $S_\alpha$ interpolates between the 
physical system for $\alpha=1$ and the reference system for $\alpha=0$. 
Integrating the relation $\partial \log Z(\alpha)/(\partial\alpha)=
-\langle \Delta S \rangle_\alpha$ I find
\be
\label{adiab}
 \log(Z(\alpha\!=\!1))=\log(Z(\alpha\!=\!0)) 
 - \int_0^1 d\alpha'\, \langle \Delta S\rangle_{\alpha'} \;\; ,
\ee
where $\langle .\rangle_\alpha$ is computed using the action $S_\alpha$. In 
the case of the anharmonic oscillator it is natural to use the harmonic 
oscillator as a reference system. In that case the reference partition
function is 
\be 
 Z(\alpha\!=\!0) = \sum_n \exp(-\beta E_n^0) 
  = \frac{\exp(-\beta\omega_0/2)}{1-\exp(-\beta\omega_0)},
\ee
where $\omega_0$ is the oscillator constant. Note that the free energy 
$F=T\log(Z)$ of the anharmonic oscillator should be independent of the 
reference frequency $\omega_0$. The integral over the coupling constant 
$\alpha$ can be calculated in a Monte Carlo simulation by slowly changing 
$\alpha$ from 0 to 1 during the simulation. Free energy calculations of 
this type play an important role in quantum chemistry, and more efficient 
methods for determining $\Delta F$ have been developed \cite{Jarzynski:1997}.

\section{Quantumchromodynamics}
\subsection{QCD at zero temperature and density}
\label{sec_qcd}

 The rich phenomenology of strong interacting matter is encoded in a 
deceptively simple Lagrangian. The fundamental fields in the Lagrangian 
are quark fields $q_{\alpha\, f}^c$ and gluon fields $A_\mu^a$. Here, $\alpha=1,
\ldots,4$ is a Dirac spinor index, $c=1,\ldots,N_c$ with $N_c=3$ is a 
color index, and $f={\it up}, {\it down},{\it strange}, {\it charm},
{\it bottom}, {\it top}$ is a flavor index. Interactions in QCD are
governed by the color degrees of freedom. The gluon field $A_\mu^a$ is 
a vector field labeled by an index $a=1,\ldots,N_c^2-1$ in the adjoint
representation. The $N_c^2-$ multiplet of gluon fields can be used to 
construct a matrix valued field $A_\mu=A_\mu^a \frac{\lambda^a}{2}$, where 
$\lambda^a$ is a set of traceless, Hermitian, $N_c\times N_c$ matrices. 
The QCD Lagrangian is
\be
\label{l_qcd}
 {\cal L } =  - \frac{1}{4} G_{\mu\nu}^a G_{\mu\nu}^a
  + \sum_f^{N_f} \bar{q}_f ( i\gamma^\mu D_\mu - m_f) q_f\, ,
\ee
where $G^a_{\mu\nu}$ is the QCD field strength tensor defined by 
\be
 G_{\mu\nu}^a = \partial_\mu A_\nu^a - \partial_\nu A_\mu^a
  + gf^{abc} A_\mu^b A_\nu^c\, ,
\ee
and $f^{abc}=4i\,{\rm Tr}([\lambda^a,\lambda^b]\lambda^c)$ are 
the $SU(N_c)$ structure constants. The action of the covariant 
derivative on the quark fields is 
\be
 i D_\mu q =  \left(
 i\partial_\mu + g A_\mu^a \frac{\lambda^a}{2}\right) q\, ,
\ee
and $m_f$ is the mass of the quarks. The terms in equ.~(\ref{l_qcd}) 
describe the interaction between quarks and gluons, as well as nonlinear 
three and four-gluon interactions. Note that, except for the number 
of flavors and their masses, the structure of the QCD Lagrangian is
completely fixed by the local $SU(N_c)$ color symmetry.

 A natural starting point for studying the phase diagram of hadronic 
matter is to consider the light flavors (up, down, and strange) as
approximately massless, and the heavy flavors (charm, bottom, top) as
infinitely massive. In this limit the QCD Lagrangian is completely
characterized by two integer valued parameters, the number of colors 
$N_c=3$ and flavors $N_f=3$, and a single dimensionless coupling 
constant $g$. Quantum fluctuations cause the coupling constant
to become scale dependent \cite{Gross:1973id,Politzer:1973fx}. At 
one-loop order the running coupling constant is
\be
\label{g_1l}
 g^2(q^2) = \frac{16\pi^2}
  {b_0\log(q^2/\Lambda_{\it QCD}^2)}\, , \hspace{1cm}
 b_0=\frac{11}{3}N_c-\frac{2}{3}N_f\, ,
\ee
where $q$ is a characteristic momentum and $N_f$ is the number of active 
flavors. The scale dependence of the coupling implies that, as a quantum 
theory, QCD is not governed by a dimensionless coupling but by a 
dimensionful scale, the QCD scale parameter $\Lambda_{\it QCD}$. This 
phenomenon is known as dimensional transmutation~\cite{Coleman:1973jx}. 

 A crucial aspect of the scale dependence of the coupling in QCD is that
the effective interaction decreases as the energy or momentum scale is 
increased. This feature of QCD is called asymptotic freedom
\cite{Gross:1973id,Politzer:1973fx}. It implies that high energy 
interactions can be analyzed using perturbative QCD. The flip side of 
asymptotic freedom is anti-screening, or confinement: The effective 
interaction between quarks increases with distance, and quarks are 
permanently confined into hadrons. The absence of colored states in 
the spectrum implies that the use of perturbation theory is subtle, 
even at high energy. Quantities that can be computed perturbatively 
either involve a sum over many hadronic states, or allow for a 
factorization of perturbative interactions and non-perturbative 
matrix elements. 

 If quarks are massless then QCD observables are dimensionless ratios
like $m_p/\Lambda_{\it QCD}$, where $m_p$ is the mass of the proton. This
implies that the QCD scale is not a parameter of the theory, but reflects
a choice of units. In the real world QCD is part of the standard model,
quarks acquire masses by electroweak symmetry breaking, and the QCD 
scale is fixed by value of the coupling constant at the weak scale. 
Experiments determine the value of the QCD fine structure constant
$\alpha_s=g^2/(4\pi)$ at the position of the $Z$ boson pole, $\alpha_s
(m_z)= 0.1184\pm 0.0007$ \cite{Nakamura:2010zzi}. The numerical value 
of $\Lambda_{QCD}$ depends on the renormalization scheme used in
computing quantum corrections to the coupling constant. Physical 
observables, as well as the value of $b_0$, are independent of this 
choice. In the modified minimal subtraction ($\overline{MS}$) scheme
the scale parameter is $\Lambda_{QCD}\simeq 200$ MeV \cite{Nakamura:2010zzi}.

\begin{figure}[t]
\bc\includegraphics[width=0.7 \textwidth]{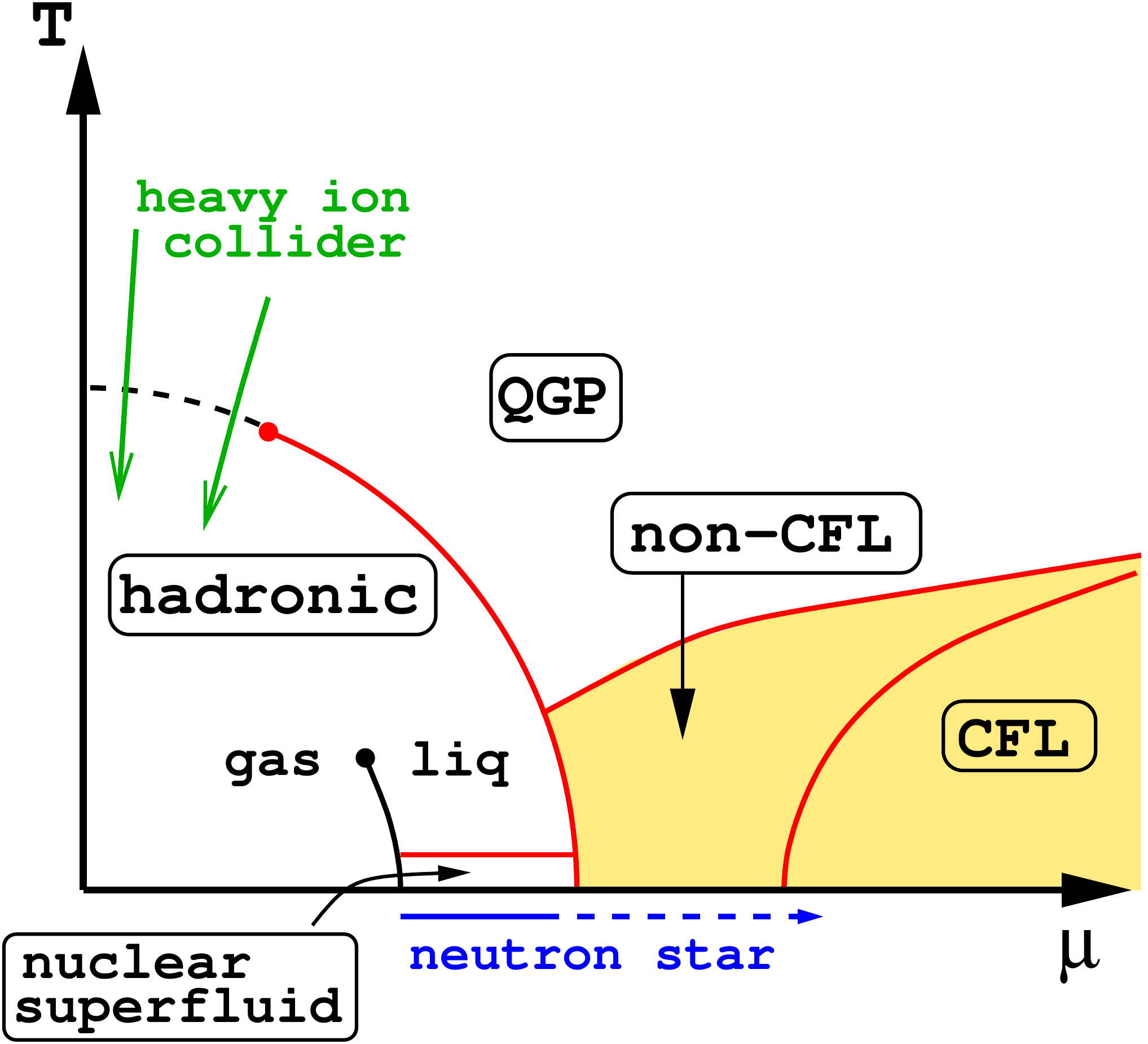}\ec
\caption{\label{fig_qcd_phase}
Schematic phase diagram of QCD as a function of temperature $T$ and
baryon chemical potential $\mu$. The quark gluon plasma phase is 
labeled QGP, and CFL refers to the color superconducting phase 
that is predicted to occur at asymptotically large chemical potential. 
The critical endpoints of the chiral and nuclear liquid-gas phase 
transitions, are denoted by red and black points, respectively. 
The chiral pseudo-critical line associated with the crossover transition
at low temperature is shown as a dashed line.  The green arrows indicate
the regions of the phase diagram that can be studied by the experimental 
heavy ion programs at RHIC and the LHC.}
\end{figure}

 A schematic phase diagram of QCD is shown in Fig.~\ref{fig_qcd_phase}.
In this figure I show the phases of strongly interacting matter as a 
function of the temperature $T$ and the baryon chemical potential $\mu$. 
The chemical potential $\mu$ controls the baryon density $\rho$, defined 
as 1/3 times the number density of quarks minus the number density of 
anti-quarks. In the following I will explain that the basic structure 
of the phase diagram is determined by asymptotic freedom and the 
symmetries of QCD. For more detailed reviews see
\cite{Alford:2007xm,Adams:2012th,Braun-Munzinger:2015hba}.

 At small temperature and chemical potential the interaction between
quarks is dominated by large distances and the effective coupling is
strong. This implies that quarks and gluons are permanently confined in
color singlet hadrons, with masses of order $\Lambda_{QCD}$. The proton,
for example, has a mass of $m_p=935$ MeV. A simplistic view of the 
structure of the proton is that it is a bound state of three constituent 
quarks with effective masses $m_Q\simeq m_p/3\simeq \Lambda_{QCD}$. These
masses should be compared to the bare up and down quark masses which are 
of the order 10 MeV.

 As a consequence of strong interactions between virtual quarks and 
anti-quarks in the QCD ground state a vacuum condensate of $\bar{q}q$ 
pairs is generated, $\langle\bar{q}q\rangle\simeq-\Lambda^3_{QCD}$
\cite{GellMann:1968rz,Coleman:1980mx,tHooft:1979bh}. This vacuum 
expectation value spontaneously breaks the approximate chiral $SU(3)_L
\times SU(3)_R$ flavor symmetry of the QCD Lagrangian down to its 
diagonal subgroup, the flavor symmetry $SU(3)_V$. Spontaneous chiral 
symmetry breaking implies the existence of Goldstone bosons, massless 
modes with the quantum numbers of the generators of the broken axial 
symmetry $SU(3)_A$. The corresponding excitations in the spectrum
of QCD are the $\pi$, $K$ and $\eta$ mesons. The $SU(3)_L\times SU(3)_R$ 
symmetry is explicitly broken by quark masses, and the mass of the 
charged pion is $m_\pi=139$ MeV. This scale can be compared to the 
mass of the lightest non-Goldstone particle, the rho meson, which 
has a mass $m_\rho=770$ MeV.

 At low energy Goldstone bosons can be described in terms of 
an effective field theory in which composite $\pi$, $K$ and 
$\eta$ particles are treated as fundamental fields. The Goldstone 
boson field can be parametrized by unitary matrices 
\be
\Sigma = \exp(i\lambda^a\phi^a/f_\pi)\, , 
\ee
where $\lambda^a$ are the Gell-Mann matrices for $SU(3)$ flavor 
and $f_\pi=93$ MeV is the pion decay constant. For example, $\pi^0=\phi^3$ 
and $\pi^\pm=(\phi_1\pm i\phi_2)/2$ describe the neutral and charged pion. 
Other components of $\phi^a$ describe the neutral and charged kaons, as 
well as the eta. The eta prime, which is the $SU(3)_F$ singlet meson,
acquires a large mass because of the axial anomaly, and is not a Goldstone 
boson. The axial anomaly refers to the fact that the flavor singlet
axial current, which is conserved in massless QCD at the classical
level, is not conserved if quantum effects are taken into account. The
divergence of the axial current $A_\mu=\bar{q}\gamma_\mu\gamma_5 q$ is 
\be 
\partial_\mu A^\mu = \frac{g^2N_f}{32\pi^2} 
\epsilon^{\mu\nu\alpha\beta}
      G^a_{\mu\nu}G^a_{\alpha\beta} \, . 
\ee
The right hand side is the topological charge density, which I will
discuss in more detail in Sect.~\ref{sec_QCD_vac}.

 At low energy the effective Lagrangian for the chiral field can be 
organized as a derivative expansion in gradients of $\Sigma$. Higher 
derivative terms describe interactions that scale as either the 
momentum or the energy of the Goldstone boson. Since Goldstone bosons 
are approximately massless, the energy is of the same order of magnitude 
as the momentum. We will see that the expansion parameter is $p/(4\pi 
f_\pi)$. At leading order in $(\partial/f_\pi)$ there is only one 
possible term which is consistent with chiral symmetry, Lorentz invariance 
and the discrete symmetries $C,P,T$. This is the Lagrangian of the 
non-linear sigma model
\be
\label{l_chpt}
{\mathcal L} = \frac{f_\pi^2}{4} {\rm Tr}\left[
 \partial_\mu\Sigma\partial^\mu\Sigma^\dagger\right] 
  +\left[ B {\rm Tr}(M\Sigma^\dagger) + h.c. \right]
+ \ldots. \, , 
\ee
where the term proportional to $B$ takes into account explicit symmetry 
breaking. Here, $M={\rm diag}(m_u,m_d,m_s)$ is the quark mass matrix 
and $B$ is a low energy constant that I will fix below. 

 First, I will show that the parameter $f_\pi$ controls the pion decay 
amplitude. For this purpose I have to gauge the weak $SU(2)_L$ symmetry
of the non-linear sigma model. 
As usual, this is achieved by promoting the derivative to a gauge covariant 
operator $\nabla_\mu\Sigma = \partial_\mu\Sigma+ig_w W_\mu\Sigma$ where 
$W_\mu$ is the charged weak gauge boson and $g_w$ is the weak coupling 
constant. The gauged non-linear sigma model gives a pion-$W$ boson 
interaction 
\be 
{\mathcal L}=g_w f_\pi W^\pm_\mu \partial^\mu \pi^\mp\, .
\ee 
This term contributes to the amplitude ${\cal A}$ for the decay $\pi^\pm\to 
W^\pm\to e^\pm\nu_e$. I get ${\cal A}=g_wf_\pi q_\mu$, where $q_\mu$ is 
the momentum of the pion. This result can be compared to the standard
definition of $f_\pi$ in terms of the weak axial current matrix element 
of the pion, $\langle 0|A_\mu^a|\pi^b\rangle = f_\pi q_\mu\delta^{ab}$. 
This comparison shows that the coefficient of the kinetic term in the
non-linear sigma model is indeed the weak decay constant of the pion.

 In the ground state $\Sigma=1$ and the ground state energy is $E_{vac}=
-2B{\rm Tr}[M]$. Using the relation $\langle\bar{q}q\rangle = \partial 
E_{vac}/(\partial m)$ we find $\langle\bar{q}q\rangle=-2B$. Fluctuations 
around $\Sigma=1$ determine the masses of the Goldstone bosons. The 
pion mass satisfies the Gell-Mann-Oaks-Renner relation (GMOR) 
\cite{GellMann:1968rz}
\be
\label{GMOR}
m_\pi^2 f_\pi^2 =-(m_u+m_d)\langle\bar{q}q\rangle
\ee
and analogous relations exist for the kaon and eta masses. This result
shows the characteristic non-analytic dependence of the pion mass on
the quark masses, $m_\pi\sim \sqrt{m_q}$.

\subsection{QCD at finite temperature}
\label{sec_qcd_T}

 The structured of QCD at high temperature can be analyzed using the 
assumption that quarks and gluons are approximately free. We will 
see that this assumption is internally consistent, and that it is 
confirmed by lattice calculations. If the temperature is large then
quarks and gluons have thermal momenta $p\sim T\gg\Lambda_{QCD}$. 
Asymptotic freedom implies that these particles are weakly interacting, 
and that they form a plasma of mobile color charges, the quark gluon 
plasma (QGP)~\cite{Shuryak:1977ut,Shuryak:1978ij}. The pressure of 
a gas of quarks and gluons is
\be  
P =\frac{\pi^2T^4}{90}\left( 2\left(N_c^2-1\right)
  + 4N_cN_f \frac{7}{8}\right)\,  . 
\ee
This is the Stefan-Boltzmann law, where $2(N_c^2-1)$ is the number
of bosonic degrees of freedom, and $4N_cN_F$ is the number of fermions. 
The factor 7/8 takes into account the difference between Bose and
Fermi statistics. The pressure of a QGP is parametrically much bigger
than the pressure of a pion gas, indicating that the QGP at high
temperature is thermodynamically stable. 

  The argument that the QGP at asymptotically high temperature is 
weakly coupled is somewhat more subtle than it might appear at first 
glance. If two quarks or gluons in the plasma interact via large angle 
scattering then the momentum transfer is large, and asymptotic freedom
implies that the effective coupling is weak. However, the color Coulomb 
interaction is dominated by small angle scattering, and it is not 
immediately clear why the effective interaction that governs small 
angle scattering is weak. The basic observation is that in a high 
temperature plasma there is a large thermal population ($n\sim T^3$) 
of mobile color charges that screen the interaction at distances beyond 
the Debye length $r_D\sim 1/(gT)$. We also note that even in the limit 
$T\gg\Lambda_{QCD}$ the QGP contains a non-perturbative sector of 
static magnetic color fields \cite{Linde:1980ts}. This sector of
the theory, corresponding to energies below the magnetic screening
scale $m_M\lsim g^2T$, is strongly coupled, but it does not contribute 
to thermodynamic or transport properties of the plasma in the limit 
$T\to\infty$.

 The quark gluon plasma exhibits neither color confinement nor chiral
symmetry breaking. This implies that the high temperature phase must
be separated from the low temperature hadronic phase by a phase transition.
The order of this transition is very sensitive to the values of the quark
masses. In QCD with massless $u,d$ and infinitely massive $s,c,b,t$ quarks
the transition is second order \cite{Pisarski:1983ms}. In the case of
massless (or sufficiently light) $u,d,s$ quarks the transition is first
order. Lattice simulations show that for realistic quark masses, $m_u
\simeq m_d\simeq 10$ MeV and $m_s\simeq 120$ MeV, the phase transition
is a rapid crossover \cite{Aoki:2006we,Bazavov:2011nk}. The transition 
temperature, defined in terms of the chiral susceptibility, is $T_c\simeq 
151\pm 3 \pm 3$ MeV \cite{Aoki:2006br,Aoki:2009sc}, which is consistent with 
the result $154 \pm 9$ MeV reported in \cite{Bazavov:2011nk,Bazavov:2014pvz}.
 
 The phase transition is expected to strengthen as a function of chemical 
potential, so that there is a critical baryon chemical potential $\mu$ at 
which the crossover turns into a first order phase transition 
\cite{Stephanov:2004wx}. This critical point is the endpoint of the chiral 
phase transition. Because of the fermion sign problem, which I will discuss 
in Sect.~\ref{sec_lQCD_mu}, it is very difficult to locate the critical 
endpoint using simulations on the lattice. Model calculations typically 
predict the existence of a critical point, but do not constrain its 
location. A number of exploratory lattice calculations have been performed 
\cite{Fodor:2001pe,Allton:2002zi,Karsch:2003va,Fodor:2004nz,Gavai:2008zr,Datta:2012pj},
but at the time I am writing these notes it has not been demonstrated
conclusively that the transition strengthens with increasing baryon 
chemical potential \cite{deForcrand:2010he}. The critical endpoint is
important because, with the exception of the endpoint of the nuclear 
liquid-gas transition, it is the only thermodynamically stable point 
in the QCD phase diagram at which the correlation length diverges. This 
means that the critical endpoint may manifest itself in heavy ion collisions 
in terms of enhanced fluctuation observables\cite{Stephanov:1998dy}.
  
\subsection{High baryon density QCD}
\label{sec_qcd_mu}
 
 The origin of the phase diagram, $T=\mu=0$, corresponds to the vacuum 
state of QCD. If we stay on the $T=0$ line and increase the chemical 
potential $\mu$ then there is no change initially. At zero temperature 
the chemical potential $\mu$ is the energy required to add a baryon to 
the system, and QCD has a large mass gap for baryonic states. The first 
non-vacuum state we encounter along the $T=0$ axis of the phase diagram 
is nuclear matter, a strongly correlated superfluid composed of approximately 
non-relativistic neutrons and protons. Nuclear matter is self-bound, and 
the baryon density changes discontinuously at the onset transition,
from $\rho=0$ to nuclear matter saturation density $\rho=\rho_0 \simeq 
0.15\,{\rm fm}^{-3}$. The discontinuity decreases as nuclear matter is 
heated, and the nuclear-liquid gas phase transition ends in a critical 
point at $T\simeq 18$ MeV and $\rho\simeq\rho_0/3$  
\cite{Sauer:1976zzf,Pochodzalla:1995xy,Elliott:2013pna}. Hot hadronic 
matter can be described quite accurately as a weakly interacting gas 
of hadronic resonances. Empirically, the density of states for both 
mesons and baryons grows exponentially. A system of this type is called 
a Hagedorn gas, and it is known that a Hagedorn gas has a limiting 
temperature. It is also known that an exponential density of states
can be realized using the string model of hadronic resonances.

 In the regime $\mu\gg\Lambda_{QCD}$ we can use arguments similar to 
those in the limit $T\gg\Lambda_{\it QCD}$ to establish that quarks and 
gluons are weakly coupled. At low temperature non-interacting quarks 
form a Fermi surface, where all states below the Fermi energy $E_F
\simeq \mu/3$ are filled, and all states above the Fermi energy 
are empty. Interactions take place near the Fermi surface, and the 
corresponding interaction is weak. The main difference between cold 
quark matter and the hot QGP is that the large density of states near 
the quark Fermi surface implies that even weak interactions can cause 
qualitative changes in the ground state of dense matter. In particular, 
attractive interactions between pairs of quarks $(\vec{p}_F,-\vec{p}_F)$
on opposite sides of the Fermi surface leads to color superconductivity 
and the formation of a $\langle qq\rangle$ diquark condensate. 

 Since quarks carry many different quantum numbers, color, flavor, and 
spin, a variety of superconducting phases are possible. The most symmetric 
of these, known as the color-flavor locked (CFL) phase, is predicted to 
exist at asymptotically high density \cite{Alford:1998mk,Schafer:1999fe}. 
In the CFL phase the diquark order parameter is 
\be 
\label{CFL}
\langle q^A_{\alpha f} q^B_{\beta g}\rangle = 
  (C\gamma_5)_{\alpha\beta} \epsilon^{ABC}\epsilon_{fgh}\delta^h_C\Phi \, , 
\ee
where $C\gamma_5$ is an anti-symmetric (spin zero) Dirac matrix, 
and $\Phi$ determines the magnitude of the gap near the Fermi surface. 
This order parameter has a number of interesting properties. It breaks 
the $U(1)$ symmetry associated with baryon number, leading to superfluidity, 
and it breaks the chiral $SU(3)_L \times SU(3)_R$ symmetry. Except for 
Goldstone modes the spectrum is fully gapped. Fermions acquire a BCS-pairing 
gap, and gauge fields are screened by the color Meissner effect. This 
implies that the CFL phase, even though it is predicted to occur in a
very dense liquid of quarks, exhibits many properties of superfluid 
nuclear matter.

 The CFL order parameter describes equal pair-condensates $\langle ud
\rangle =\langle us\rangle = \langle ds\rangle$  of all three light quark 
flavors. As the density is lowered effects of the non-zero strange quark 
mass become important, and less symmetric phases are predicted to appear 
\cite{Alford:2007xm}. Phases that have been theoretically explored include 
Bose condensates of pions and kaons, hyperon matter, states with inhomogeneous 
quark-anti-quark or diquark condensates, and less symmetric color 
superconducting phases. The regimes of moderate baryon chemical potential 
in the phase diagram shown in Fig.~\ref{fig_qcd_phase} is largely conjecture. 
Empirical evidence shows that at low $\mu$ there is a nuclear matter phase 
with broken chiral symmetry and zero strangeness, and weak coupling 
calculations indicate that at high $\mu$ we find the CFL phase with 
broken chiral symmetry but non-zero strangeness. In principle the two 
phases could be separated by a single onset transition for strangeness
\cite{Schafer:1998ef,Hatsuda:2006ps}, but model calculation support a 
richer picture in which one or more first order transitions intervene, 
as indicated in Fig.~\ref{fig_qcd_phase}. 

\section{Lattice QCD}
\label{sec_lqcd}
\subsection{The Wilson action}
\label{sec_wilson}

 Symmetry arguments and perturbative calculations can be used to 
establish general features of the QCD phase diagram, but quantitative 
results can only be obtained using numerical calculations based on 
lattice QCD. The same is true for the masses of hadrons, their properties, 
and interactions. Lattice QCD is based on the euclidean path integral 
representation of the partition function, see the contribution by Hatsuda
and \cite{Creutz:1983,Montvay:1994,Smit:2002,Gattringer:2009,Lin:2014} for 
introductions. More detailed reviews of the lattice field theory approach 
to hot and dense QCD can be found in \cite{Fodor:2009ax,Ding:2015ona}. 

 The euclidean partition function for QCD is 
\be
 Z(T,\mu,V) = \int {\cal D}A_\mu\, {\cal D}q_f\, {\cal D}\bar{q}_f
 \; \exp(-S_E) \, , 
\ee
where $S_E$ is the euclidean action 
\be 
 S_E = -\int_0^\beta d\tau \int_V d^3x\; {\cal L}^E\, , 
\ee
$\beta=T^{-1}$ is the inverse temperature and ${\cal L}^E$ is the 
euclidean Lagrangian, which is obtained by analytically continuing 
equ.~(\ref{l_qcd}) to imaginary time $\tau=it$. As in the quantum
mechanical example in equ.~(\ref{z}) the temperature enters via
the boundary condition on the fields in the imaginary time direction. 
Gauge fields and fermions obey periodic and anti-periodic boundary 
conditions, respectively. The chemical potential enters through its
coupling to the conserved baryon density  
\be 
 {\cal L}^E(\mu) =  {\cal L}^E(0) + \mu \bar{q}_f\gamma_0 q_f\, . 
\ee
In his pioneering work Wilson proposed to discretize the action
on a $N_\tau\times N_\sigma^3$ space-time lattice with lattice spacings
$a_\tau$ and $a_\sigma$ \cite{Wilson:1974sk}. In many cases $a_\sigma=
a_\tau=a$, but we will encounter an exception in Sect.~\ref{sec_cl_QCD}.
when we discuss the Hamiltonian formulation of the theory.

 At finite temperature we have to ensure that the spatial volume 
is larger than the inverse temperature, $L>\beta$. Here, $\beta=N_\tau 
a_\tau$, $L=N_\sigma a_\sigma$, and $V=L^3$ is the volume.  Thermodynamic 
quantities are determined by taking derivatives of the partition
function. The energy and baryon density are given by 
\begin{eqnarray}
\label{e_part}
{\cal E} &=& -\frac{1}{V} \left.\frac{\partial\log Z}{\partial\beta}
  \right|_{\beta\mu}\, , \\
\label{n_part}
 \rho  &=& \;\frac{1}{\beta V} \left.\frac{\partial\log Z}{\partial\mu}
  \right|_{\beta}\, .
\end{eqnarray}
The discretized action for the gauge fields originally suggested by 
Wilson is given by 
\be 
\label{s_wilson}
S_W = - \frac{2}{g^2}\sum_n\sum_{\mu<\nu} {\rm Re}\,{\rm Tr} 
  \left[ W_{\mu\nu}(n) -1 \right]
\ee
where $W_{\mu\nu}(n)$ is the plaquette, the product of gauge links around
an elementary loop on the lattice, 
\be
\label{plaq}
 W_{\mu\nu}(n) = U_\mu(n)U_\nu(n+\hat\mu)U_{-\mu}(n+\hat\mu+\hat\nu)
                U_{-\nu}(n+\hat{\nu})\, . 
\ee
Here, $n=(n_\tau,n_i)$ labels lattice sites and $\hat\mu$ is a unit
vector in the $\mu$-direction. The gauge links $U_\mu(n)$ are $SU(N_c)$ 
matrices. We can think of the gauge links as line integrals 
\be 
\label{link_var}
U_\mu(n)=\exp(ia A_\mu(n))\, , 
\ee
and of the plaquettes as fluxes
\be 
 W_{\mu\nu}(n))=\exp(ia^2 G_{\mu\nu}(n))\, ,
\ee
but the fundamental variables in the path integral are the (compact) 
group variables $U_\mu$, not the (non-compact) gauge potentials $A_\mu$.
In particular, the path integral in pure gauge QCD takes the form
\be 
\label{Z_Wilson}
 Z = \int \prod_{n,\mu} dU_\mu(n)\, \exp(-S_W)\, ,
\ee
where $dU$ is the Haar measure on $SU(N_c)$. The Haar measure describes
the correct integration measure for the gauge group. Some group integrals
are discussed by Hatsuda, but part of the beauty of the Metropolis method
is that we never have to explicitly construct $dU_\mu(n)$. 

Using equ.~(\ref{link_var})
we can check that the Wilson action reduces to continuum pure gauge
theory in the limit $a\to 0$. We note that the gauge invariance of QCD 
is maintained exactly, even on a finite lattice, but that Lorentz invariance 
is only restored in the continuum limit. We also observe that classical scale 
invariance implies that the massless QCD action is independent of $a$. The 
continuum limit is taken by adjusting the bare coupling at the scale of 
the lattice spacing according to asymptotic freedom, see equ.~(\ref{g_1l}). 
In practice the lattice spacing is not small enough to ensure the accuracy 
of this method, and more sophisticated scale setting procedures are 
used \cite{Fodor:2009ax,Ding:2015ona}.

 Monte Carlo simulations of the path integral equ.~(\ref{Z_Wilson}) can 
be performed using the Metropolis algorithm  explained in 
Sect.~\ref{sec_qm}:

\begin{itemize}
\item Initialize the link variables with random
$SU(N_c)$ matrices. A simple algorithm is based on writing $U$
in terms of $N_c$ complex row vectors $\vec{u}_i$. Take each vector
to be random unit vector and then use the Gram-Schmidt method to 
orthogonalize the different vectors, $\vec{u}_i\cdot\vec{u}_j^*
=\delta_{ij}$. This ensures that $U$ is unitary and distributed
according to the $SU(N_c)$ Haar measure \cite{Mezzadri:2006}.

\item Sweep through the lattice and update individual link variables. 
For this purpose multiply the link variable by a random $SU(N_c)$
matrix, $U_\mu\to R U_\mu$. Compute the change in the Wilson action 
and accept the update with probability $\exp(-\Delta S_W)$. 

\item Compute physical observables. The simplest observable is the 
average plaquette $\langle W_{\mu\nu}\rangle$, which can be related
to the equation of state, see equ.~(\ref{e_part}). More complicated
observables include the correlation function between plaquettes, and 
the Wilson loop
\be 
W({\cal C}) = Tr\left[ L({\cal C})\right] \, , 
\hspace{0.5cm}
L({\cal C})= \prod_{(n,\mu)\in{\cal C}} U_\mu(n)\, , 
\ee
where $L({\cal C})$ is the product of link variables around a closed
loop. The average Wilson loop is related to the potential between
two static charges in the fundamental representation
\be 
 V(R) = -\lim_{T\to\infty}\frac{1}{T} 
    \log \left[\langle W({\cal C}) \rangle \right]\, ,
\ee
where $R\times T$ is the area of a rectangular loop ${\cal C}$. 

\item Tune to the continuum limit $a\to 0$ by adjusting the coupling 
constant according to the asymptotic freedom formula equ.~(\ref{g_1l}). 
Note that the Lambda parameter for the lattice regulator is quite small, 
$\Lambda_{\it lat} = \Lambda_{\bar{MS}}/28.8$ \cite{Hasenfratz:1980kn}. 
Also observe that we have to increase $N_\sigma,N_\tau$ to keep the 
physical volume constant as $a\to 0$. Indeed, once the continuum limit
$a\to 0$ is reached we have to study the infinite volume (thermodynamic)
limit $V\to \infty$. This is more difficult than it appears, because 
$a\to 0$, corresponding to $g\to 0$, is a critical point of the 
partition function (\ref{Z_Wilson}), and simulations exhibit critical 
slowing down. 

\end{itemize}

Metropolis simulations with the pure gauge Wilson action are very simple 
and robust. As an illustration I provide a simple $Z_2$ lattice gauge 
theory code written by M.~Creutz in the appendix. Reasonable results for the 
heavy quark potential can be obtained on fairly coarse lattices, for example 
an $8^4$ lattice with a spacing $a\simeq 0.25$ fm \cite{Lepage:1998dt}.
However, accurate results with controlled error bars require significant 
computational resources. In practice the perturbative relation between 
$a$ and $g^2$ is only valid on very fine lattices, and the scale setting 
has to be done non-perturbatively. Also, determining the spectrum of pure 
gauge theory is difficult. Purely gluonic states, glueballs, are quite heavy, 
with masses in the range $m\simeq 1.6$ GeV and higher. This implies that 
gluonic correlation functions are short range, requiring a resolution 
$a\simeq 0.1$ fm or better. Finally, simulations on fine lattices are 
affected by critical slowing down. Indeed, finding an efficient method 
for updating gauge fields on very fine lattices, analogous to the cluster 
algorithms for spin models \cite{Wolff:1988uh}, is an important unsolved 
problem.

\subsection{Fermions on the lattice}
\label{sec_fermions}
 
 The main difficulty in lattice QCD is related to the presence
of light fermions. The fermion action is of the form
\be 
 S_F= a^4 \sum_{m,n} \bar{q}(m)D_{mn}q(n)\, . 
\ee
Formally, the integration over the fermion fields 
can be performed exactly, resulting in the determinant of the Dirac 
operator $\det(D(A_\mu,\mu))$. Several methods exist for discretizing 
the Dirac operator $D$, and for sampling the determinant. Different 
discretization schemes differ in the degree to which chiral symmetry 
is maintained on a finite lattice. The original formulation due to Wilson 
\cite{Wilson:1974sk} preserves no chiral symmetry, the staggered Fermion 
scheme \cite{Kogut:1974ag} maintains a subset of the full chiral symmetry, 
while the domain wall \cite{Kaplan:1992bt} and overlap methods
\cite{Neuberger:1997fp} aim to preserve the full chiral symmetry on 
a discrete lattice. 

 The central difficulty in implementing these methods is that the 
fermion determinant is a very non-local object. While updating a
single gauge link only requires recalculating a small number of 
plaquettes (6 in $d=4$ dimensions) in the Wilson action, recalculating
the fermion action requires computing the determinant of a (very sparse) 
matrix of size $(N_\tau N_\sigma^3)\times(N_\tau N_\sigma^3)$ or larger. 
This is clearly impractical. Fermion algorithms rely on a number of 
tricks. The first is the observation that the Dirac operator has a 
property called $\gamma_5$-hermiticity, $\gamma_5 D\gamma_5=D^\dagger$,
which implies that $\det(D)$ is real. The determinant of a two-flavor 
theory is then real and positive. This allows us to rewrite the 
fermion determinant as a path integral over a bosonic field with a 
non-local but positive action
\be 
\det(D_u)\det(D_d) = 
\det(DD^\dagger) = \int {\cal D}\phi{\cal D}\phi^\dagger\, 
  \exp(-\phi^\dagger (DD^\dagger)^{-1}\phi) \, . 
\ee
The path integral over the pseudofermion field $\phi$ can be sampled 
using a combination of deterministic methods like molecular dynamics 
and stochastic methods such as the Metropolis algorithm. These combined 
algorithms are known as Hybrid Monte Carlo (HMC) methods. Codes that
implement the HMC algorithm for pseudofermions are significantly
more complicated than the Metropolis algorithm for the pure gauge
Wilson action discussed above, and I refer the interested reader 
to the more specialized literature \cite{Luscher:2010ae}. I also
note that since these algorithms involve the calculation of $D^{-1}$
the computational cost increases as the quark masses are lowered. 

 The calculation of correlation functions also differs from the 
bosonic case. Consider, for example, an operator with the quantum
numbers of a charged pion, $J_\pi(x)=\bar{u}^a(x)\gamma_5 d^a(x)$. Since 
the fermion action is quadratic the correlation function in a given
gauge configuration can be computed exactly in terms of the fermion
propagator. The full correlation function is 
\be
 \Pi_\pi(x) = \langle J_\pi(x)J_\pi(0)\rangle = 
  \left\langle {\rm Tr}\left[ S(x,0)\gamma_5 S(0,x)\gamma_5\right]
  \right\rangle\, , 
\ee
where $S(x,y)=\langle x|D^{-1}|y\rangle$ is the fermion propagator, and
we have assumed exact isospin symmetry so that the propagator of the 
up quark is equal to the propagator of the down quark. Note that the
interaction between quarks is encoded in the average over all gauge 
fields. The one-gluon exchange interaction, for example, corresponds
to a perturbative fluctuation in the gauge field that modifies the 
two quark propagators. An operator with the quantum number of the 
proton is $\eta_\alpha(x) =\epsilon_{abc}(u^a(x)C\gamma_\mu u^b(x))(
\gamma^\mu\gamma_5 d^c(x))_\alpha$. The correlation function is
\be 
 \Pi_{\alpha\beta}(x) = 2\epsilon_{abc}\epsilon_{a'b'c'}
   \Big\langle 
   \left(\gamma_\mu\gamma_5 S^{cc'}(0,x)\gamma_\nu\gamma_5\right)_{\alpha\beta}
   {\rm Tr}\left[\gamma_\mu S^{aa'}(0,x)\gamma_\nu C(S^{bb'}(0,x))^TC\right]
   \Big\rangle \, . 
\ee
Note that meson correlation function involves one forward and one 
backward going propagator, whereas the propagators in the baryon 
correlation function are all forward going. A difficulty arises when
we consider flavor singlet $\bar{q}{q}$ currents such as $J_{\eta'}=
(\bar{u}^a(x)\gamma_5 u^a(x)+\bar{d}^a(x)\gamma_5 d^a(x))/\sqrt{2}$,
which has the quantum numbers of the $\eta'$ meson. We find 
\be
 \Pi_{\eta'}(x) = \langle J_{\eta'}(x)J_{\eta'}(0)\rangle = 
  \left\langle {\rm Tr}\left[ S(x,0)\gamma_5 S(0,x)\gamma_5\right]
     -2{\rm Tr}\left[ S(x,x)\gamma_5\right]
      {\rm Tr}\left[ S(0,0)\gamma_5\right]
  \right\rangle\, , 
\ee
which involve propagators $S(x,x)$ that loop back to the same point.
These contributions are known as quark-line disconnected diagrams, 
and difficult to treat numerically, see \cite{Endress:2014qpa} for
a recent discussion.

\begin{figure}[t]
\bc\includegraphics[width=0.7\textwidth]{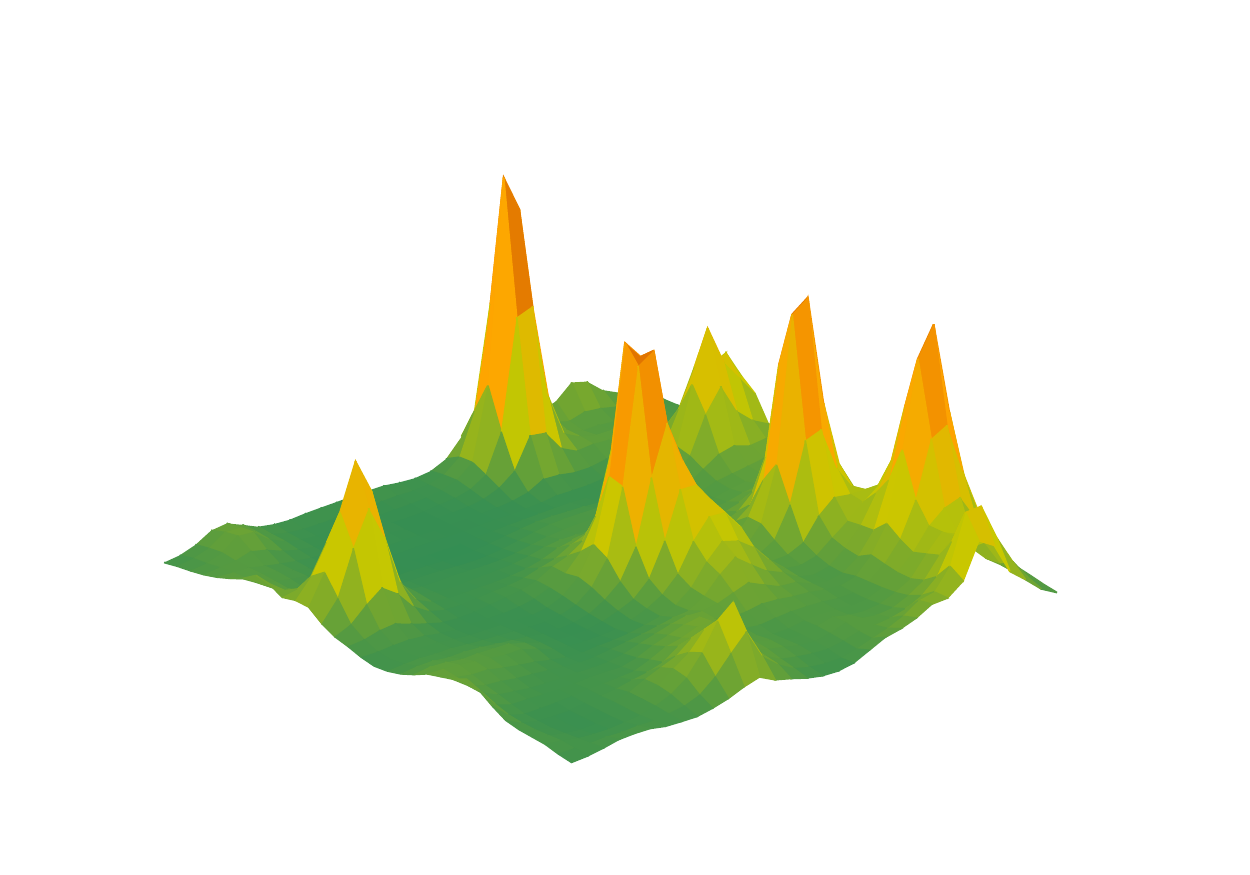}\ec
\caption{\label{fig_top}
Topological objects in lattice QCD (figure courtesy of S.~Sharma, 
see \cite{Dick:2015twa}). This picture shows a slice through a 
low lying eigenstate of the Dirac operator in lattice QCD. }
\end{figure}

\subsection{The QCD vacuum}
\label{sec_QCD_vac}

  It is natural to hope that lattice QCD can provide us with an intuitive 
picture of what the QCD vacuum looks like, similar to the picture of the 
quantum mechanical ground state shown in Fig.~\ref{fig_path}. This turns 
out to be more complicated, for a number of reasons. The first is that the 
field in QCD is a $SU(3)$ matrix, which is hard to visualize. The second, 
more important, problem is related to quantum fluctuations. In QCD there
is no obvious separation of scales that would allow us to clearly 
separate perturbative fluctuations from large semi-classical fluctuations.

 This has led to the idea to eliminate short range fluctuations by 
some kind of filtering or smoothing algorithm. The simplest of these is
known as cooling \cite{Teper:1985ek}. In the cooling method we modify
the Metropolis algorithm so that only updates that reduce the action
are accepted. Since the update algorithm is local, this will tend to
eliminate small structures but preserve larger objects. A modern version
of cooling is gradient flow \cite{Luscher:2010iy}. In the gradient 
flow method we continue the gauge fields to a 5th ``time'' dimension. In 
this direction the fields satisfy a differential equation
\be 
\label{grad_flow} 
 \partial_\tau A_\mu = D^\nu G_{\mu\nu}\, ,
\ee
where $A_\mu(\tau=0)$ is the four-dimensional gauge field and the rhs
is computed from the gauge potentials evaluated at the flow time $\tau$.
The Lorentz indices remain four-dimensional. The rhs of the flow equations
is the classical equation of motion, so that the gradient flow tends to
drive gauge fields towards the closest classical solution. The only 
finite action solutions of the euclidean field equations on $R^4$ are
instantons \cite{Belavin:1975fg,Schafer:1996wv}. Instantons and 
anti-instantons are characterized by integer values $Q_{\it top}=\pm 1$
of the topological charge 
\be 
\label{q_top}
 Q_{\it top} = \int d^4x \, q(x)\, , \hspace{0.5cm}
 q(x) = \frac{g^2}{64\pi^2} \epsilon^{\mu\nu\alpha\beta}
        G^a_{\mu\nu}G^a_{\alpha\beta}\, . 
\ee
Exact higher charge solutions exist, but the QCD vacuum is dominated
by configurations with both instantons and anti-instantons. These
gauge field configurations are only approximate solutions of the equations 
of motion \cite{Schafer:1996wv}. Under cooling or gradient flow instantons 
and anti-instantons will eventually annihilate and evolve to an exact 
multi-instantons solution with $Q_{\it top}=N_I-N_A$ , where $N_{I,A}$ are 
the numbers of (anti)instantons. However, the $N_I+N_A$ topological objects 
are preserved for flow times that are much longer than the decay time of 
ordinary quantum fluctuations, and the total number of well separated 
instantons and anti-instantons can be determined. 

 The average topological charge is zero, but the pure gauge vacuum 
is characterized by a non-zero topological susceptibility
\be
 \chi_{\it top} = \frac{1}{V} \langle Q_{\it top}^2\rangle \, , 
\ee
where $V$ is the euclidean four-volume. The topological charge
can be determined using the naive lattice discretization of 
equ.~(\ref{q_top}), but this operator is very noisy, and in general not 
an integer. This problem can be addressed using the cooling or gradient 
flow algorithms discussed above. Recent lattice calculations based 
on these methods give $\chi_{\it top} = (190\pm 5\,{\rm MeV})^4$ 
\cite{DelDebbio:2004ns,Ce:2014sfa}. A simple picture of the QCD vacuum 
which is consistent with this value is the dilute instanton liquid
model, which assumes that the topological susceptibility is determined
by Poisson fluctuations in an ensemble of instantons and anti-instantons 
with an average density $(N_I+N_A)/V\simeq 1 \, {\rm fm}^{-4}$
\cite{Schafer:1996wv}. This is an approximate picture, and more 
complicated configurations involving monopoles and fractional charges
are needed to understand the large $N_c$ limit and the role of 
confinement \cite{Poppitz:2012nz}. 

 Another important development is the use of fermionic methods
to analyze the vacuum structure of QCD. In a given gauge configuration
the quark propagator can written as 
\be 
 S(x,y) = \sum_\lambda \frac{\psi_\lambda(x)\psi^\dagger_\lambda(y)}
  {\lambda+im}\, ,
\ee
where $\psi_\lambda$ is an eigenvector of the Dirac operator with 
eigenvalue $\lambda$: $D\psi_\lambda= (\lambda+im)\psi_\lambda$. Note that 
this is not how propagators are typically determined in lattice QCD,
because the calculation of the complete spectrum is numerically very 
expensive. Gamma five hermiticity implies that eigenvalues come 
in pairs $\pm \lambda$. The quark condensate is given by
\be 
 \langle \bar{q}q \rangle = -i\int d^4x\, 
   \left\langle {\rm Tr}\left[S(x,x)\right]\right\rangle 
 = -\left\langle \sum_{\lambda > 0} \frac{2m}{\lambda^2+m^2}
      \right\rangle\, .
\ee
Here, I have ignored the contribution from exact zero modes because
the density of zero modes is suppressed by $m^{N_f}$. This factor
comes from the determinant in the measure. If we were to ignore the
determinant (this is called the quenched approximation), then the 
quark condensate would diverge as $1/m$. We observe that a finite 
value of the quark condensate in the chiral limit $m\to 0$ requires
an accumulation of eigenvalues near zero. This can be made more explicit 
by introducing the density of states
\be 
 \rho(\nu) = \left\langle \sum_{\lambda\geq 0}\delta(\lambda-\nu) 
   \right\rangle \, .   
\ee
The chiral condensate in the thermodynamic and chiral limits is 
given by 
\be 
 \langle\bar{q}q\rangle  = - \pi\rho(0)\, . 
\ee
This is known as the Banks-Casher relation \cite{Banks:1979yr}.
Note that is is essential to take the thermodynamic $V\to\infty$ limit 
before the chiral limit $m\to 0$. 

 Exact zero modes of the Dirac operator are related to topology. The 
Dirac operator has one left handed zero mode in the field of an instanton, 
and a right handed zero mode in the field of an anti-instanton. This 
is consistent with the Atiyah-Singer index theorem, which states that 
the topological charge is equal to the index of the Dirac operator, the 
difference between the number of left and right handed zero modes,
$Q_{\it top}=N_f(n_L-n_R)$. These results suggest that it is possible 
to give a purely fermionic definition of the topological charge density. 

 On the lattice, this can be achieved for a class of Dirac operators
that satisfy the Ginsparg-Wilson relation \cite{Ginsparg:1981bj}
\be 
 D\gamma_5 + \gamma_5 D = a D\gamma_5 D \, , 
\ee
where $a$ is the lattice spacing. In the continuum limit we recover
the expected relation $ D\gamma_5 + \gamma_5 D =0$ for the massless
Dirac operator. The important observation is that the fermionic 
topological density
\be 
\label{q_ferm}
q_f(n) = \frac{1}{2a^3}\, {\rm tr}_{CD}\left[ \gamma_5 D(n,n)\right] \, ,
\ee
where ${\rm tr}_{CD}$ is a color-Dirac trace, satisfies the index
theorem 
\be 
 Q_{\it top}=a^4\sum_n \; q_f(n)\,  
\ee
on a discrete lattice. 
Fig.~\ref{fig_top} shows the absolute square of $q_f(x)$ constructed 
from lying eigenmodes of the Dirac operator. We observe that 
fermionic operators can indeed be used to probe the topological content 
of the QCD vacuum directly, without the need for filtering or smoothing.

 The existence of zero mode implies that the topological susceptibility 
is zero if at least one quark flavor is massless. This is because the
path integral measure contains the fermion determinant, which is 
vanishes if $m=0$ and $Q_{\it top}\neq 0$. We can be more precise 
using the chiral lagrangian equ.~(\ref{l_chpt}). In order to keep 
track of topology we add to the QCD action a topological term 
$S_\theta = i\theta Q_{\it top}$. Then the topological susceptibility 
is given by the second derivative of the free energy with respect 
to $\theta$. Since every zero mode in the Dirac operator contributes 
a factor ${\rm det}(M)$ to the partition function we know that 
$\theta$ enters the effective lagrangian in the combination $\theta
+{\rm arg}(\det(M))$. The vacuum energy is determined by 
\be 
\label{v_theta}
 V =-B {\rm Tr}\left[Me^{i\theta/N_f}\Sigma^\dagger\right] + {\it h.c.} \, , 
\ee
and we observe that the topological susceptibility in QCD with degenerate
quark masses is proportional to $m\langle\bar{q}q\rangle$. Note that 
equ.~(\ref{v_theta}) is consistent with the vanishing of $\chi_{\it top}$ 
for $m_u=0$. If $m_u=0$ and $m_d\neq 0$ then equ.~(\ref{v_theta}) is 
minimized by $\Sigma=\exp(i\phi\tau_3)$ with $\phi=\theta/2$, and the 
vacuum energy is independent of $\theta$.   

 It is tempting to think that exact zero modes, governed by topology, 
and approximate zero modes, connected to chiral symmetry breaking,
are related. This is the basis of the instanton liquid model
\cite{Schafer:1996wv}. In the instanton liquid model we consider an 
ensemble of instantons and anti-instantons with no (or small) net 
topology. The exact zero modes of individual instantons are lifted, 
and form a zero mode zone. The density of eigenvalues in the zero 
mode zone determines the chiral condensate via the Banks-Casher relation. 
This model predicts the correct order of magnitude of $\langle\bar{q}q
\rangle$, but the calculation cannot be systematically improved
because chiral symmetry breaking requires strong coupling. Recently,
we showed that the connection of chiral symmetry breaking, instantons
and monopoles can be made precise in a certain limit of QCD. The
idea is to compactify QCD on $R^3\times S_1$, where the size of the 
circle is much smaller than $\Lambda^{-1}_{\it QCD}$, and the fermions 
satisfy non-thermal (twisted) boundary conditions \cite{Cherman:2016hcd}.

\subsection{Lattice QCD at finite baryon density}
\label{sec_lQCD_mu}
 
 In section \ref{sec_fermions} I discussed some of the difficulties 
that appear when we discretize the Dirac operator.  A separate, more 
serious, issue with fermions is that for $\mu\neq 0$ the Dirac
operator does not satisfy $\gamma_5$-hermiticity. This implies that  
the fermion determinant is no longer real, and that standard importance 
sampling methods fail. This is the ``sign'' problem already mentioned in 
Sect.~\ref{sec_qcd_T}. To understand the severity of the problem 
consider a generic expectation value 
\be
\langle {\cal O} \rangle = 
  \frac{\int dU \, \det(D)\,{\cal O}\,e^{-S}}
       {\int dU \, \det(D)\,e^{-S}}\, . 
\ee
If the determinant is complex I can write this as
\be
\label{O_pq}
\langle {\cal O} \rangle = 
  \frac{\int dU \, |\det(D)|\,{\cal O}e^{i\varphi}\,e^{-S}}
       {\int dU \, |\det(D)|\,e^{i\varphi}e^{-S}}
 \equiv \frac{\langle {\cal O}e^{i\varphi}\rangle_{pq}}
             {\langle e^{i\varphi}\rangle_{pq}}\, , 
\ee
where $\langle .\rangle_{pq}$ refers to a phase quenched average. This
average can be computed using the Metropolis (or HMC) algorithm. The 
problem is that the average phase $\langle e^{i\varphi}\rangle_{pq}$ is
very small. This follows from the fact that the average phase can
be expressed as the ratio of two partition functions
\be
\label{ph_av}
\langle e^{i\varphi} \rangle_{pq} = 
  \frac{\int dU \, \det(D)\,\,e^{-S}}
       {\int dU \, |\det(D)|\,e^{-S}}
 = \frac{Z}{Z_{pq}} = e^{-V\Delta F} \, , 
\ee
where $\Delta F$ is the free energy density difference, and $V$ is the 
volume of the system. This shows that the phase is exponentially small, 
and that the ratio equ.~(\ref{O_pq}) is very difficult to compute. 

As a specific example consider QCD with two degenerate flavors, up and
down, and a baryon chemical potential $\mu_u=\mu_d=\mu_B/3$. Then 
$\det(D)=\det(D_u)\det(D_d)$ and $|\det(D)|=\det(D_u)\det(D_d)^*$.
The phase quenched partition function 
$Z_{pq}$ can be interpreted as the partition function of QCD with a
non-zero isospin chemical potential $\mu_u=-\mu_d=\mu_I/2$. The small
$\mu$ behavior of both the isospin and baryon number theories at $T=0$
is easily understood. The isospin theory has a second order phase
transition at $\mu_I=m_\pi$ which corresponds to the onset of pion
condensation. The baryon theory has a first order transition at
$\mu_B=m_p-B$, where $B\simeq 15$ MeV is the binding energy of infinite
nuclear matter. This implies that for $\mu>m_\pi$ the partition functions
$Z$ and $Z_{pq}$ describe very different physical systems, and the 
sign problem is severe. 

 The sign problem may manifest itself in different ways. Consider,
for example, an attempt to study the correlation function of $A$ 
nucleons in a QCD ensemble generated at $\mu_B=0$. For large $A$ 
this correlation function determines the binding energy of nuclear 
matter. There are two difficulties with this approach. The first is 
that the operator contains $3A$ quark fields, so that the correlator
has up to $(3A)!$ contractions. This is not prohibitive, because
the number of contractions can be reduced using symmetries and 
iterative algorithms. Indeed, correlators for medium mass nuclei 
have been computed \cite{Lin:2014}. The second, more serious, problem
is the signal-to-noise ratio. The variance of the correlator $C$ is 
\be 
 {\rm var}(C)= \langle CC^\dagger \rangle - \langle C\rangle^2\, . 
\ee
The $A$ nucleon correlator $C$ contains $3A$ forward going quark propagators,
and $CC^\dagger$ consists of $3A$ forward and $3A$ backward propagators. This
implies that $CC^\dagger$ couples to a state of $3A$ mesons. Since the lightest
meson is the pion and the lightest baryon is the proton the signal-to-noise
of an $A$ nucleon correlation function is 
\be 
 \frac{\cal S}{\cal N} \sim \exp(-A(m_p -3m_\pi/2)\tau) \, . 
\ee
In order to resolve the ground state with a given $A$ we have to 
make $\tau$ sufficiently large so that excited states with the same
$A$ are suppressed. For $A=1$ there is a $\pi N$ continuum starting 
an excitation energy $\Delta E=m_\pi$, and the first resonance at 
$\Delta E=m_\Delta-m_N\simeq 300$ MeV. This means that we have to 
consider $\tau\gsim 1$ fm. For multi-nucleon states the situation
is more complicated, because there are many closely spaced multi-nucleon
states in a finite volume. The problem is studied, for example, in 
\cite{Beane:2003da}. The conclusion is that different bound and 
scattering states are separated by 10s of MeV, requiring $\tau\gsim
4$ fm. It may be possible to improve on this estimate by using 
variationally improved sources, but even for $\tau\simeq 2$ fm the 
signal to noise is extremely poor for $A\gsim 4$. This shows that in 
simulations with fixed $A$ the sign problem manifests itself as a 
noise problem. This is not surprising. One way to think about the 
sign problem is to view it as an overlap problem. The configurations
that contribute to $Z_{pq}$ have poor overlap with those that contribute
to $Z$. The same phenomenon is at work here. Configurations generated
at $\mu_B=0$ reflect vacuum physics, and the lightest fermionic 
fluctuation is a pion. Large cancellations are required to explore the
physics of multi-baryon states.

 There are many attempts to find direct solutions to the sign problem, 
but at this time the only regime in which controlled calculations
are feasible is the regime of small $\mu$ and high $T$. In this region
the partition function can be expanded in a Taylor series in $\mu/T$. 
The corresponding expansion coefficients are generalized susceptibilities
that can be determined from lattice simulations at zero chemical
potential. The susceptibilities not only determine the equation of state
at finite baryon density, but also control fluctuations of conserved
charges. 

In addition to methods that are restricted to the regime $\mu
\lsim \pi T$, a number of proposals to explore QCD at high baryon 
density are being pursued. This includes new approaches, like integration 
over Lefshetz thimbles \cite{Cristoforetti:2012su,Aarts:2014nxa},
as well as novel variants of old approaches, like the complex Langevin
method \cite{Aarts:2009uq,Sexty:2013ica}, or the use of dual variables 
\cite{Kloiber:2013rba}. The ultimate promise of these methods is still 
unclear, but the central importance of the sign problem to computational 
physics continues to attract new ideas.

\subsection{Real time properties}
\label{sec_lQCD_real}

 The basic trick in lattice QCD is the continuation of the path 
integral to imaginary time. This makes it possible to calculate 
the path integral by importance sampling, but it implies that we 
only have direct access to imaginary time correlation functions. 
For many observables this is not a serious problem. Thermodynamic
observables, for example, are static quantities and no analytic 
continuation is necessary. The ground state contribution to a hadron 
correlation function is $\Pi(\tau)\sim e^{-m_H\tau}$ which is trivially 
continued to $\Pi(t)\sim e^{-im_Ht}$. However, difficulties arise if one 
studies excited states, in particular resonances, the interaction between
hadrons, or the real time response of many body systems at finite 
temperature and density. 

 Significant progress has been made in studying scattering processes,
at least in the elastic regime. This is discussed in some of the later 
chapters of this book. Here, I will concentrate on the calculation
of real time response functions. The prototypical example is the 
calculation of the shear viscosity of a QCD plasma using the retarded 
correlation function of the stress tensor $T^{xy}$, 
\be
\label{G_ret}
G^{xy,xy}_R(\omega,\vec{k}) = -i\int dt\int d^3x\, 
  e^{i(\omega t-\vec{k}\cdot\vec{x})} \Theta(t)
  \langle \left[ T^{xy}(\vec{x},t),T^{xy}(0,0)\right]\rangle\, , 
\ee
The associated spectral function is defined by $\rho(\omega,\vec{k})
=-\,{\rm Im}\,G_R(\omega,\vec{k})$. The imaginary part of the retarded 
correlator is a measure of dissipation. This relation can be made 
more precise using fluid dynamics, which is an effective theory of 
the response function in the low energy, small momentum limit 
\cite{Schafer:2009dj,Schaefer:2014awa}. 

 Linearized fluid dynamics shows that the static response function is
determined by the pressure of the fluid, and that the leading energy and
momentum dependence is governed by transport coefficients. These
relations can be used to derive Kubo formulas, expressions for the 
transport coefficients in terms of retarded correlation functions. 
The Kubo relation for the shear viscosity is 
\be 
\label{eta_kubo}
\eta = \lim_{\omega\to 0} \lim_{k\to 0} 
   \frac{\rho^{xy,xy}(\omega,\vec{k})}{\omega}\, ,
\ee
and similar results hold for the bulk viscosity, the thermal conductivity, 
and heavy quark diffusion constants. 

 The spectral function contains information about the physical
excitations that carry the response. The euclidean path integral does 
not provide direct access to the retarded correlator or the spectral
function. Lattice calculations are based on the relation between the 
spectral function and the imaginary energy (Matsubara) correlation 
function 
\be 
\label{G_E_w}
G_E(i\omega_n)= \int \frac{d\omega}{2\pi} \frac{\rho(\omega)}
 {\omega-i\omega_n}\, , 
\ee
where $\omega_n=2\pi nT$ is the Matsubara frequency. The imaginary
time correlation function is 
\be 
\label{G_E_tau}
G_E(\tau)= \int \frac{d\omega}{2\pi} K(\omega,\tau) \rho(\omega) \, , 
\ee
where the kernel $K(\omega,\tau)$ is given by 
\be 
\label{Ker}
K(\omega,\tau) = \frac{\cosh[\omega(\tau-1/(2T))]}{\sinh[\omega/(2T)]}
 =  \left[1+n_B(\omega)\right] e^{-\omega\tau}
      + n_B(\omega)e^{\omega\tau}\, ,
\ee
and $n_B(\omega)$ is the Bose distribution function. The imaginary time 
correlation function equ.~(\ref{G_E_tau}) was studied on the lattice 
in \cite{Karsch:1986cq,Meyer:2007ic,Meyer:2007dy,Sakai:2007cm}. The 
basic idea for calculating transport coefficients is to numerically 
compute $G_E(\tau)$, invert the integral transform in equ.~(\ref{G_E_tau}) 
to obtain the spectral functions $\rho(\omega)$, and then study the 
limit $\omega\to 0$.

 The problem is that $G_E(\tau)$ is computed on a small number of 
discrete lattice sites, and that the imaginary time correlator 
at distances on the order of $\beta/2$ is not very sensitive to the 
slope of the spectral function at low energy. Recent attempts to 
to address these problems and to obtain numerically stable spectral 
functions and reliable error estimates are based on Bayesian methods
such as the maximum entropy method mentioned in Sect.~\ref{sec_qm}, 
see \cite{Aarts:2007wj,Aarts:2007va}. It is also possible to 
optimize the contribution from the transport peak by measuring 
the correlation functions of conserved charges, such as energy and 
momentum density, at non-zero spatial momentum 
\cite{Aarts:2006wt,Meyer:2008gt}. A possible issue with lattice 
calculations is that effects of poor resolution tend to bias the result
towards small values of $\eta/s$, where $s$ is the entropy density. 
The finite temperature spectral function satisfies 
the sum rule \cite{Romatschke:2009ng}
\be 
\frac{2}{\pi} \int d\omega\, 
  \left[ \eta(\omega) -\eta_{T\!=\!0}(\omega)\right] = 
  \frac{3}{10}\, sT\, , 
\ee
where $\eta(\omega) = \rho(\omega)/\omega$. On the lattice it is 
difficult to resolve sharp features in the spectral function. Roughly,
the resolution is limited by the lowest Matsubara frequency $\pi T$. I 
will therefore assume that the $T\neq 0$ spectral function is a Lorentzian
with width $\pi T$
\be
\eta(\omega) -\eta_{T\!=\!0}(\omega) \simeq
 \frac{\eta(0)(\pi T)^2}{\omega^2 + (\pi T)^2}\, . 
\ee
Then the integral on the lhs is equal to $\eta(0)\pi T$, and the sum
rule predicts $\eta/s\sim 3/(10\pi)$, quite close to $\eta/s=1/(4\pi)$. 
The lesson is that it is easy to obtain small values of $\eta/s$, and 
much more difficult to obtain large values of $\eta/s$, predicted by 
perturbative QCD \cite{Arnold:2000dr}. 

 The first calculation of the shear viscosity on the lattice was 
performed by Karsch and Wyld \cite{Karsch:1986cq}. More recently, the 
problem of computing the shear and and bulk viscosity in a pure gauge
plasma near $T_c$ was revisited by Meyer \cite{Meyer:2007ic,Meyer:2008gt}. 
He obtains $\eta/s=0.102(56)$ and $\zeta/s=0.065(17)$ at $T=1.24T_c$. Shear 
viscosity is only weakly dependent on temperature, but bulk viscosity is
strongly peaked near $T_c$. The value of $\eta/s$ is consistent with 
experimental results, and with the prediction from holographic duality,
$\eta/s=1/(4\pi)$ \cite{Kovtun:2004de}.

\begin{figure}[t]
\bc\includegraphics[width=0.95 \textwidth]{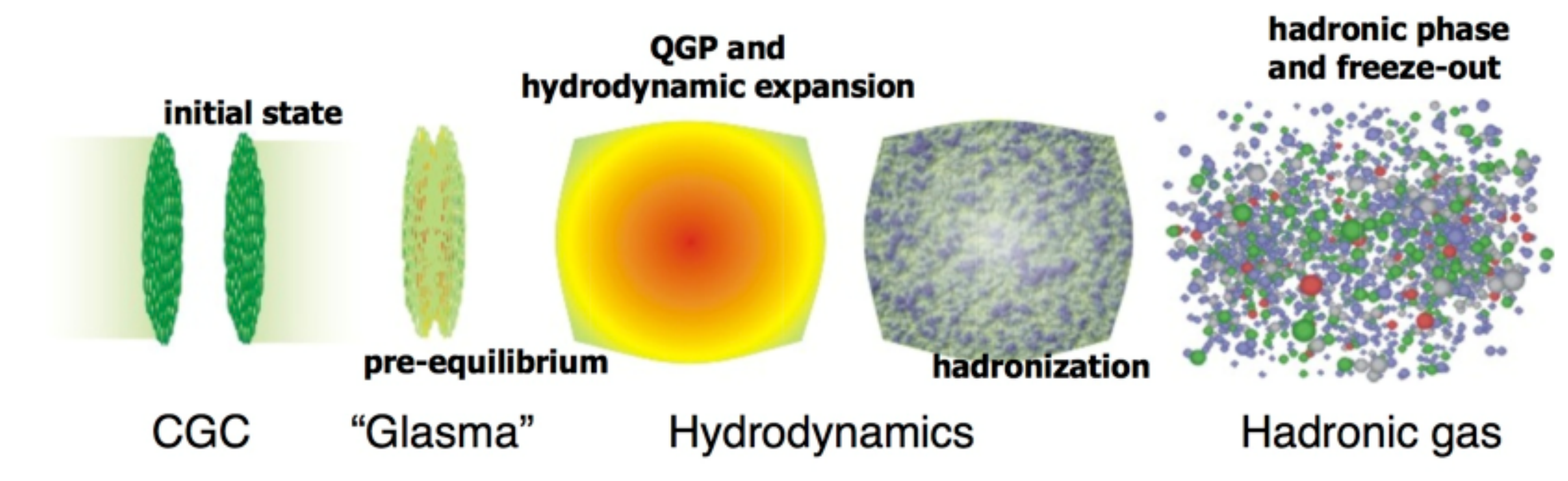}\ec
\caption{\label{fig_hic}
Schematic time evolution of a heavy ion collision. Figure courtesy 
of S.~Bass. CGC refers to the color glass condensate, a semi-classical
model of the overpopulated gluon configuration in the initial state
of a heavy ion collision. Glasma refers to the non-equilibrium evolution 
of this state into a locally equilibrated plasma. Hydrodynamics is 
the theory of the time evolution of a locally equilibrated fireball,
and hadronic phase refers to the late time kinetic stage of the
collision. }
\end{figure}

\section{Nonequilibrium QCD}
\label{sec_neq_qcd}

 In the remainder of this chapter I will discuss a number of coarse
grained approaches to the non-equilibrium dynamics of QCD. These 
method are relevant to the study of nuclear collisions, in particular
in the ultra-relativistic regime. This regime is explored experimentally
at the Relativistic Heavy Ion Collider (RHIC) at Brookhaven National
Laboratory and the Large Hadron Collider (LHC) at CERN. A rough 
time line of a heavy ion collision is shown in Fig.~\ref{fig_hic}.
Initial nucleon-nucleon collisions release a large number of quarks 
and gluons. This process is described by the full non-equilibrium
quantum field theory, but there are a number of approximate descriptions 
that may be useful in certain regimes. The first is a classical 
field theory description in terms of highly occupied classical 
gluon fields. The second is a kinetic theory in terms of quark and
gluon quasi-particles. Finally, there is a new approach, which is a 
description in terms of a dual gravitational theory. 

 Theories of the initial state demonstrate that there is a tendency
towards local equilibration. If local equilibrium is achieved then a 
simpler theory, fluid dynamics is applicable. Fluid dynamics is 
very efficient in the sense that it deals with a small number of 
variables, the conserved densities of particle number, energy and 
momentum, and that it has very few parameters, an equation of state 
and a set of transport coefficients. The fluid dynamic stage of a 
heavy ion collision has a finite duration. Eventually the density
becomes too low and local equilibrium can no longer be maintained. 
At this point kinetic theory is again relevant, now formulated in
terms of hadronic quasi-particles. All the theories we have 
mentioned, fluid dynamics, kinetic theory, classical field theory, 
and holography, have reached a high degree of sophistication and
I will point to text books and review for detailed introductions.
Nevertheless, the basic ideas are quite simple, and I will provide
some examples in the following sections.

\subsection{Fluid Dynamics}
\label{sec_hydro}

 I begin with fluid dynamics, because it is the most general and in some 
ways the simplest non-equilibrium theory. It is important to remember, 
however, that fluid dynamics is a very rich framework, both mathematically 
and in terms of the range of phenomena that one may encounter. In the 
following I will focus on the non-relativistic theory. There is no 
fundamental difference between the relativistic and non-relativistic 
theories, but some simplifications appear in the non-relativistic regime. 
Non-relativistic fluid dynamics is used in many areas of physics, including 
the physics of cold atomic Fermi gases and neutron stars. The relativistic 
theory is relevant to high energy heavy ion collisions and supernova 
explosions. Introductions to relativistic fluid dynamics can be found in
\cite{Romatschke:2009im,Rezzolla:2013,Jeon:2015dfa}.

 Fluid dynamics reduces the complicated non-equilibrium many-body
problem to equations of motion for the conserved charges. The 
reason that this is possible is the separation of scales between
the microscopic collision time $\tau_{\it micro}$, and the relaxation
time $\tau_{\it macro}$ of hydrodynamic variables. A generic perturbation
of the system decays on a time scale on the order of $\tau_{\it micro}$,
irrespective of the typical length scale involved. Here, $\tau_{\it micro}$
is determined by microscopic time scales, such as the typical collision
time between quasi-particles. A fluctuation of a conserved charge, on 
the other hand, cannot decay locally and has to relax by diffusion or 
propagation. The relevant time scale $\tau_{\it macro}$ increases with 
the length scale of the perturbation. As a consequence, when we focus 
on sufficiently large scales we can assume $\tau_{\it macro}\gg 
\tau_{\it micro}$, and focus on the evolution of conserved charges. 

 In a simple non-relativistic fluid the conserved charges are the mass 
density $\rho$, the momentum density $\vec{\pi}$, and the energy density 
${\cal E}$. The momentum density can be used to define the fluid velocity, 
$\vec{u}=\vec{\pi}/\rho$. By Galilean invariance the energy density can 
then be written as the sum of the internal energy density and the 
kinetic energy density, ${\cal E}={\cal E}_0+\frac{1}{2}\rho u^2$. The 
conservation laws are
\bea
\label{hydro1}
\frac{\partial \rho}{\partial t} &=& 
   - \vec{\nabla}\cdot\vec{\pi}   , \\
\label{hydro2}
 \frac{\partial \pi_i}{\partial t} &=&  
   - \nabla_j\Pi_{ij}, \\
\label{hydro3}
 \frac{\partial {\cal E}}{\partial t} &=&
   - \vec{\nabla} \cdot\vec{j}^{\epsilon} .  
\eea 
In order for these equations to close we have to specify constitutive 
relations for the stress tensor $\Pi_{ij}$ and the energy current 
$\vec{j}^{\epsilon}$. Since fluid dynamics is an effective long
wavelength theory we expect that the currents can be systematically
expanded in gradients of the hydrodynamic variables $\rho$, $\vec{u}$
and ${\cal E}_0$. At leading order the stress tensor contains no derivatives
and the structure is completely fixed by rotational symmetry and
Galilean invariance. We have 
\be 
 \Pi_{ij} = \rho u_i u_j + P\delta_{ij}+ \delta \Pi_{ij}\, ,
\ee
where $P=P(\rho,{\cal E}_0)$ is the equation of state and $\delta\Pi_{ij}$ 
contains gradient terms. The approximation $\delta\Pi_{ij}=0$ is 
called ideal fluid dynamics, and the equation of motion for $\vec\pi$ 
is known as the Euler equation. Ideal fluid dynamics is time reversal 
invariant and the entropy is conserved. If gradient terms are included
then time reversal invariance is broken and the entropy increases.
We will refer to  $\delta\Pi_{ij}$ as the dissipative stresses. At
first order  in the gradient expansion  $\delta\Pi_{ij}$ can be 
written as $\delta\Pi_{ij}=-\eta\sigma_{ij}-\zeta\delta_{ij}\langle 
\sigma\rangle$ with
\be 
 \sigma_{ij} = \nabla_i u_j +\nabla_j u_i 
  -\frac{2}{3}\delta_{ij}   \langle\sigma\rangle \, ,
\hspace{0.1\hsize}
 \langle\sigma\rangle =\vec{\nabla}\cdot\vec{u}\, . 
\ee
The dissipative stresses are determined by two transport coefficients, 
the shear viscosity $\eta$ and the bulk viscosity $\zeta$. The energy 
current is given by 
\be
\vec{j}^{\;\epsilon} = \vec{u}w+\delta\vec{j}^{\epsilon}\, , 
\ee
where $w=P+{\cal E}$ is the enthalpy. At leading order in the gradient 
expansion 
\be
\label{j_kappa}
\delta j_i^{\epsilon}=u_j\delta\Pi_{ij}-\kappa\nabla_i T\, , 
\ee
where $\kappa$ is the thermal conductivity. The second law of thermodynamics 
implies that $\eta,\zeta$ and $\kappa$ must be positive. The equation of 
motion for $\vec{\pi}$ at first order in gradients is known as the 
Navier-Stokes equation, and equ.~(\ref{j_kappa}) is Fourier's law
of heat conduction. 

 It is sometimes useful to rewrite the fluid dynamic equations using 
the comoving derivatives $D_t=\partial_t +\vec{u}\cdot\vec{\nabla}$. The
equations are
\bea
\label{rho_lag}
 D_t\rho &=& -\rho \vec{\nabla}\cdot \vec{u}\, , \\
\label{u_lag}
 D_t u_i & = & - \frac{1}{\rho} \nabla_j\left( \delta_{ij} P 
 + \delta \Pi_{ij} \right) \, , \\
\label{e_lag}
 D_t \epsilon & = & - \frac{1}{\rho} \nabla_i \left( u_i P 
 + \delta j^{\cal E}_i \right) \, ,
\eea
where $\epsilon={\cal E}/\rho$ is the energy per mass. This is called 
the Lagrangian form of the equations, in contrast to the Eulerian form 
given above. The Eulerian form is more naturally implemented on a fixed 
space-time lattice, whereas the Lagrangian form lends itself to a 
discretization where the computational cell is dragged along with the 
fluid.

\subsection{Computational fluid dynamics}
\label{sec_hydro_num}

The fluid dynamic equations form a set of partial differential 
equations (PDEs) that can be solved in a variety of ways. I will 
focus here on grid based methods. The main difficulties that a numerical 
method needs to address are: i) The existence of surfaces of 
discontinuity (shocks), ii) the need to implement global conservation
laws exactly, even on a coarse lattice, iii) the existence of 
instabilities (turbulence), and the need to deal with solutions
that involve many different length scales. 

 In the following I will discuss a numerical scheme that addresses
these issues in a fairly efficient way, the PPM algorithm of Collela 
and Woodward \cite{Colella:1984}, as implemented in the VH1 code by 
Blondin and Lufkin \cite{Blondin:1993} and extended to viscous fluids
in \cite{Schafer:2010dv}. The first observation is that it is sufficient
to construct a 1-d algorithm. Higher dimensional methods can be constructed
by combining updates in different directions. Note that the coordinate
system can be curvilinear, for example 3-d spherical or cylindrical
coordinates, or the Milne coordinate system that is used for 
longitudinally expanding quark gluon plasmas. 

 The basic 1-d algorithm consists of a Lagrangian time step followed
by a remap onto an Eulerian grid. I will denote the 1-d velocity by
$u$ and write the equation of mass conservation in terms of a mass
variable $m$
\be 
\label{hydro_1_m}
 \frac{\partial\tau}{\partial t}- \frac{\partial u}{\partial m}=0\, , 
\ee
where $\tau=\rho^{-1}$ and 
\be 
\label{m_def}
 m(r) = \int_{r_0}^{r}dr\,  \rho(r)\, .
\ee
Here, I restrict myself to flat coordinate systems. In curvilinear
coordinates equ.~(\ref{hydro_1_m}) and (\ref{m_def}) contain suitable
volume factors. Equ.~(\ref{hydro_1_m}) is solved by 
\be 
\label{r_lag}
 \frac{dr}{dt}= u\left(m(r),t\right)\, , 
\ee
which is the equation for the Lagrange coordinate. In terms of the mass
coordinate $m(r)$ the momentum and energy equations are 
\bea
\label{hydro_2_m}
\frac{\partial u}{\partial t} + \frac{\partial P}{\partial m} &=& 0 \, , \\
\label{hydro_3_m}
\frac{\partial \epsilon}{\partial t} 
  + \frac{\partial (uP)}{\partial m} &=& 0\, , 
\eea
where I have only written down the ideal contributions to the stress
tensor and energy current. To put these equations on a grid I focus
on the mass integrated quantities
\be 
U_j^n = \frac{1}{\Delta m_j} \int_{m_{j-1/2}}^{m_{j+1/2}} U(m,t^n) dm
\ee
where $U$ is any of the hydrodynamic variables $(\tau,u,\epsilon)$, 
$\Delta m_j$ is the mass contained in the cell $j$, and $m_{j+1/2}=
\sum_{k}^{j}\Delta m_k$. We can now integrate the conservation laws
(\ref{hydro_2_m},\ref{hydro_1_m}). The result is 
\bea
u_j^{n+1} &=& u_j^n + \frac{\Delta t}{\Delta m_j} 
 \left( \bar{P}_{j-1/2}-\bar{P}_{j+1/2}\right)\, ,  \\
\epsilon_j^{n+1} &=& \epsilon_j^n + \frac{\Delta t}{\Delta m_j} 
 \left( \bar{u}_{j-1/2}\bar{P}_{j-1/2}-\bar{u}_{j+1/2}\bar{P}_{j+1/2}\right) \, ,
\eea
where I have introduced the cell face averages $\bar{u}_{j\pm 1/2}$ and
$\bar{P}_{j\pm 1/2}$. These quantities can be obtained by parabolic 
interpolation from the cell integrated values. The PPM scheme
introduced in \cite{Colella:1984} uses a method for constructing 
cell face averages which conserves the cell integrated variables. 

 This scheme addresses the second issue mentioned above. The first
issue, the existence of shocks, can be taken into account by refining 
the method for calculating the cell face averages. The observation is 
that one can make use of exact solution of the equations of fluid dynamics 
in the case of piecewise constant one-dimensional flows, known as the 
Riemann problem. We can view $\bar{u}_{j+1/2}$ and $\bar{P}_{j+1/2}$ as
the solution of a Riemann problem with left state $u_j,P_j$ and 
right state $u_{j+1},P_{j+1}$. The PPM code contains a simple iterative 
Riemann solver described in \cite{Colella:1984}. Using $\bar{u}_{j\pm 1/2}$ 
and $\bar{P}_{j\pm 1/2}$ the Lagrange step is given by:

\vspace*{0.3cm} 
\begin{lstlisting}
do n = nmin-3, nmax+3

! density evolution. lagrangian code, so all we have to do is watch the 
! change in the geometry.

  r(n) = r(n) * ( dvol1(n) / dvol(n) )
  r(n) = max(r(n),smallr)

! velocity evolution due to pressure acceleration and forces.

  uold (n) = u(n)
  u(n) = u(n) - dtbdm(n)*(pmid(n+1)-pmid(n))*0.5*(amid(n+1)+amid(n)) &
        + 0.5*dt*(fict0(n)+fict1(n)) 

! total energy evolution

  e(n) = e(n) - dtbdm(n)*(amid(n+1)*upmid(n+1) - amid(n)*upmid(n))  
  q(n) = e(n) - 0.5*(u(n)**2+v(n)**2+w(n)**2)
  p(n) = max(r(n)*q(n)*gamm,smallp)

enddo
\end{lstlisting}

\vspace*{0.3cm} 
Here, ${\tt r(n)}$ is the density, ${\tt u(n)}$ is the velocity, 
and ${\tt e(n)}$ is the energy per mass. The transverse components
of the velocity are ${\tt v(n),w(n)}$. In cartesian coordinates
the volume and area factors ${\tt dvol(n),amid(n)}$ are equal to unity, 
and  the fictitious forces ${\tt fict(n)}$ vanish.

 After the Lagrange step the hydrodynamic variables have to be remapped
onto a fixed Eulerian grid. This can be done using the parabolic interpolation 
mentioned above. The advantage of the remap step is that it is simple to 
change the grid resolution in the process. Finally, we have to specify the 
time step and grid resolution. The grid resolution is determined by the 
requirement that $(\Delta x)\nabla_x U\ll U$, where $\Delta x$ is the cell 
size, and $U$ is any of the hydrodynamic variables. Note that there is 
no need to worry about discontinuities, because shocks are captured by 
the Riemann solver. Also note that the PPM scheme has at least second 
order accuracy, so that relatively coarse grids can be used. The time 
step is determined by the Courant criterion $c\Delta x \leq \Delta t$, 
where $c$ is the maximum of the speed of sound and the local flow velocity. 
This criterion ensures that the domain of dependence of any physical
variable does not exceed the cell size. 

 In general, modern hydro codes are very fast and efficient. The main 
difficulty is that $3+1$ dimensional simulations may require a lot 
of memory, and that some physical phenomena, such as turbulent 
convection and shock instabilities in supernovae, require very high 
resolution. One of the frontiers of numerical hydrodynamics is the 
problem of dealing with systems that transition from fluid dynamics
to ballistic behavior at either early or late times, or systems in
which the density varies by a very large factor. Problems of this 
type arise in the early and late time dynamics of heavy ion collisions, 
the dilute corona of cold atomic gases, and the transition from 
hydrodynamics to free streaming in the neutrino transport in a 
supernova explosions. Recent progress in this direction includes
the development of the anisotropic hydrodynamics method 
\cite{Florkowski:2010cf,Martinez:2010sc,Bluhm:2015raa,Bluhm:2015bzi}, 
and applications of the lattice Boltzmann method to problems in nuclear 
and atomic physics \cite{Romatschke:2011hm,Brewer:2015hua}.

 In the relativistic regime recent progress includes the development of 
stable and causal viscous fluid dynamics codes 
\cite{Romatschke:2009im,Jeon:2015dfa}. The problem with a naive 
implementation of the relativistic Navier-Stokes equation derived
by Landau is that viscous stresses are determined by the instantaneous
value of the shear strain $\nabla_i u_j$. This leads to acausal
propagation of shear waves and possible instabilities. This is not 
a fundamental problem with fluid dynamics. Acausal behavior occurs 
in the regime of high wave numbers in which fluid dynamics is not 
expected to be reliable. However, high wave number instabilities 
prohibit numerical implementations. The solution is to go to next order
in the gradient expansion, which includes the finite relaxation time 
of viscous stresses. In practice, second order fluid dynamics codes
are usually based on the idea of transient fluid dynamics. In this
method, the shear stresses $\delta\Pi_{ij}$ are promoted to fluid
dynamic variables, which satisfy separate fluid dynamic equations, 
see \cite{Romatschke:2009im,Jeon:2015dfa}.

\subsection{Kinetic theory}

 Fluid dynamics is based on the assumption of local thermal equilibrium
and requires the mean free path to be small compared to the characteristic
scales of the problem. When this condition is not satisfied a more 
microscopic approach to the non-equilibrium problem is required. The
simplest method of this type is kinetic theory, which is based on the 
existence of well defined quasi-particles. This implies, in particular,
that the width of a quasi-particle has to be small compared to its energy. 
In this case we can define the phase space density $f(\vec{x},\vec{p},t)$
of quasi-particles. In general, there can be many different kinds of 
quasi-particles, labeled by their spin, charge, and other quantum numbers. 
The phase space distribution determines the conserved densities that enter
the hydrodynamic description. For example, the mass density is given by
\be 
 \rho(\vec{x},t) = \int d\Gamma\, m f(\vec{x},\vec{p},t) \, ,
\ee
where $d\Gamma=d^3p/(2\pi)^3$. The momentum density is
\be 
 \vec{\pi}(\vec{x},t) = \int d\Gamma\, m v_p f(\vec{x},\vec{p},t) \, ,
\ee
where $v_p=\nabla_p E_p$ is the quasi-particle velocity and $E_p$ is 
the quasi-particle energy. In general, the quasi-particle energy can
be a functional of the phase distribution $f(\vec{x},\vec{p},t)$. This 
takes into account possible in-medium modifications of particle 
properties. If $E_p$ is a functional of $f(\vec{x},\vec{p},t)$ then
the total energy of the system is not just given by the integral of
$E_p f(\vec{x},\vec{p},t)$. Instead, we must construct an energy 
density functional ${\cal E}[f]$ that satisfies \cite{Kadanoff}
\be 
 E_p = \frac{\delta {\cal E}}{\delta f_p}\, .
\ee
The equation of motion for the distribution function is the Boltzmann
equation
\be
\label{be}
\left( \partial_t + \vec{v}\cdot\vec{\nabla}_x 
                  - \vec{F}\cdot\vec{\nabla}_p \right) 
  f(\vec{x},\vec{p},t) = C[f]\, , 
\ee
where $\vec{F}=-\vec\nabla_x E_p$ is a force, and $C[f_p]$ is the collision 
term. For dilute systems the collision term is dominated by binary scattering
and 
\be 
 C[f_p] = -\prod_{i=2,3,4}\Big(\int d\Gamma_{i}\Big) w(1,2;3,4)
   \left( f_1f_2-f_3f_4\right)\, , 
\ee
where $f_i=f(\vec{x},\vec{p}_i,t)$. The transition rate is given by
\be
w(1,2;3,4) = (2\pi)^4\delta\Big(\sum_i E_i\Big)
         \delta\Big(\sum_i \vec{p}_i\Big) \,|{\cal A}|^2\, ,
\ee  
where ${\cal A}$ is the scattering amplitude. For non-relativistic $s$-wave 
scattering ${\cal A} = 4\pi a/m$, where $a$ is the scattering length. 

 The Boltzmann equation is a 6+1 dimensional partial integro-differential
equation, and direct methods of integration, similar to those used 
in computational fluid dynamics, are impractical. Standard methods for
solving the Boltzmann equation rely on sampling phase space using 
Monte Carlo methods. In nuclear physics the test particle method for
solving the Boltzmann equation was popularized by Bertsch and Das Gupta
\cite{Bertsch:1988ik}. Below, I will present a simple non-relativistic 
algorithm described by Lepers et al.~\cite{Lepers:2010be}.
 
 The main idea is to represent the distribution as a sum of delta
functions
\be
\label{test_part}
f(\vec{x},\vec{p},t)= \frac{N}{N_t} \sum_{i=1}^{N_t}
    (2\pi)^3 \delta(\vec{p}-\vec{p}_i(t))\delta(\vec{x}-\vec{x}_i(t)) \,,
\ee
where $N$ is the number of particles, the integral of $f(\vec{x},\vec{p},t)$
over phase space, and $N_t$ is the number of test particles. In typical
applications $N_t\gg N$, but if $N$ is already very large it is possible
to run simulations with $N_t<N$. Phase space averages can be computed
as averages over test particles
\be
\bar{F}  = \frac{1}{N}\int d^3x\int d\Gamma\, f(\vec{x},\vec{p},t)
 F(\vec{x},\vec{p}) 
 =  \frac{1}{N_t}\sum_{i=1}^{N_t} F(\vec{x}_i,\vec{p}_i)\,.
\ee
In practice this requires some smoothing, and the delta functions
are replaced by Gaussian distributions
\be
\label{gauss_sm}
\delta(\vec{p}-\vec{p}_i) \delta(\vec{x}-\vec{x}_i)
 \to  g_{w_p}(\vec{p}-\vec{p}_i)g_{w_x}(\vec{x}-\vec{x}_i)\, ,
\ee
where $g_w(\vec{x})$ is a normalized Gaussian with width $w$. The 
widths $w_x$ and $w_p$ are chosen such that the delta function
singularities are smoothed out, but physical structures of the 
distribution function $f(\vec{x},\vec{p},t)$ are preserved. 

 If there is no collision term the equation of motion for the 
distribution function is Hamilton's equation for the test particle
positions and momenta
\be
 \frac{d\vec{x}_i}{dt} = \frac{\vec{p}_i}{m} \, , \hspace{1cm}
 \frac{d\vec{p}_i}{dt} = \vec{F}_i\, . 
\ee
These equations can be solved with high accuracy using a staggered
leapfrog algorithm
\bea
\label{leap_1}
\vec{v}_i(t_{n+1/2}) &=& \vec{v}_i(t_n) + \vec{a}_i(t_n) \Delta t/2\, ,\\
\label{leap_2}
\vec{r}_i(t_{n+1})  &=& \vec{r}_i(t_n)  + \vec{v}_i(t_{n+1/2})\Delta t\, ,\\
\label{leap_3}
\vec{v}_i(t_{n+1})  &=& \vec{v}_i(t_{n+1/2}) + \vec{a}_i(t_{n+1})\Delta t/2\, ,
\eea
where $\vec{a}_i=\vec{F}_i/m$ is the acceleration of particle $i$, and 
$\Delta t = t_{n+1}-t_n$ is the time step of the algorithm. The size of 
the time step depends on the specific problem, but a good check is 
provided by monitoring conservation of energy. 

 The collision term is treated stochastically, by allowing the test particles
to collide with the scaled cross section $\sigma_t = (N/N_t)\sigma$. To 
determine whether a collision occurs we go through all pairs of particles
and compute the relative distance $\vec{r}_{ij} = \vec{r}_i-\vec{r}_j$
and velocity $\vec{v}_{ij} =\vec{v}_i-\vec{v}_j$. We then determine whether 
on the current trajectory the time of closest approach will be reached during 
the next time step. This happens if $t_{\it min} = t_n-\vec{r}_{ij}\cdot
\vec{v}_{ij}/\vec{v}_{ij}^2$ satisfies $|t_{\it min}-t_n|\leq \Delta t/2$. 
In that case we compute  
\be
r_{\it min}^2 =\vec{r}_{ij}^2-
 \frac{(\vec{r}_{ij}\cdot \vec{v}_{ij})^2}{\vec{v}_{ij}^2} 
\ee
and check if $\pi r_{\it min}^2 < \sigma_t$. If this condition is satisfied 
then the collision is allowed to take place. For an $s$-wave elastic 
collision we propagate the particles to $t_{\it min}$, randomize their 
relative velocity $\vec{v}_{ij}$, and then propagate them back to $t_n$. 
Higher partial wave amplitudes are easy to implement by randomizing 
$\vec{v}_{ij}$ with suitable probability distributions. After all pairs 
have been checked we perform the velocity and position update in 
equ.~(\ref{leap_1}-\ref{leap_3}). 

 There are a number of refinements that can be included. At low temperature 
Pauli-blocking has to be taken into account. This can be done by computing
the phase space densities $f(\vec{r}_i,\vec{p}_i,t)$ for the collision
products, and accepting the collision with probability $(1-f_i)(1-f_j)$.
At higher energies relativistic effects are important. Relativistic 
effects in the particle propagation are easy to incorporate, but the
treatment of the collision term is more subtle. The problem is that 
a finite collision cross section, treated geometrically, will lead
to instantaneous interactions at a distance. Additional difficulties
arise from the treatment of resonances, pair production and annihilation, 
$n$-body processes, etc. There are a number of codes on the market 
that address these issues, and that have been tuned against existing
data on $pp$, $pA$ and $AA$ interactions in the relativistic 
regime. Examples include UrQMD \cite{Bass:1998ca}, GiBUU \cite{Buss:2011mx}, 
HSD \cite{Ehehalt:1996uq}, and others. 

 At high energies the initial $pp$ collisions are very inelastic, and
one has to rely on Monte Carlo generators developed in the high energy
physics community. A possible alternative is to use a purely partonic
kinetic theory that involves scattering between quark and gluon 
quasi-particles. There are some subtleties with this approach, having 
to do with the problem of how to include screening and damping of the 
exchanged gluons, soft gluon radiation, etc. I will not attempt to 
discuss these issues here, and I refer the reader to the original 
literature \cite{Geiger:1991nj,Xu:2004mz}.

\subsection{Classical field theory}
\label{sec_cl_QCD}

 An interesting simplification occurs if the occupation numbers are large,
$f\gg 1$. This is argued to happen for the gluons in the initial state of a 
heavy ion collision \cite{McLerran:1993ni}. In this limit the classical 
kinetic theory is equivalent to a classical field theory \cite{Mueller:2002gd}.
Indeed, if the occupations numbers are non-perturbative, $f\gsim 1/g$, the 
kinetic theory no longer applies, and we have to rely on classical field 
theory. In general the classical action is not known, but in the weak 
coupling limit the bare QCD action can be used. 
 
 Classical QCD simulation have been used to study a number of issues,
such as particle production from an overpopulated gluon field, and 
the possible approach to thermal equilibrium. Instabilities in the 
classical field evolution may play an important role in speeding
up the equilibration process. Here, I will briefly describe a
method for solving classical evolution equations on a space-time
lattice, following the recent review \cite{Mrowczynski:2016etf}.

 In order to construct a Hamiltonian approach to lattice QCD I 
start from the Wilson action in Minkowski space with separate 
coupling constants $\beta_0$ and $\beta_s$ in the temporal and
spatial direction  
\be
\label{s_lat_aniso}
S[U]=-\frac{\beta_0}{2N_c}\sum_x \sum_{i=1}^3
   {\rm Tr}\left( W_{0i}(x) + W^\dagger_{0i}(x) - 2\right)
+ \frac{\beta_s}{2N_c}\sum_x \sum_{i<j} 
   {\rm Tr}\left( W_{ij}(x)+ W^\dagger_{ij}(x)- 2\right) ,
\ee
In the continuum limit, we expect
\be 
\beta_0 = \frac{2N_c a}{g^2\Delta t}\, , \hspace{0.5cm}
\beta_s = \frac{2N_c\Delta t}{g^2a}\, . 
\ee
where $a$ and $\Delta t$ are spatial and temporal lattice spacings. 
In order to construct a Hamiltonian we have to fix the gauge freedom
of the theory. Here, I will use the temporal axial gauge, $A_0=0$. 
In this case the canonical variables are the spatial gauge potentials
and the conjugate momenta are the electric fields. On the lattice
the gauge $A_0=0$ corresponds to setting all temporal gauge links 
to the identity, $U_0(x)=1$. The canonical variables are given by
the spatial gauge links $U_j(x)$, and the conjugate momenta are 
the  temporal plaquettes $W_{0j}(x)$. In the continuum limit 
\bea
\label{A_j_latt}
A_j^a(x)  &=& \frac{2i}{ag} \, {\rm Tr} \big[\lambda^a U_j(x) \big]  , \\
\label{E_j_latt}
E^a_j(x)  &=& \frac{2i}{ag\Delta t} \, {\rm Tr} \big[\lambda^a W_{0j}(x) \big] .
\eea
Varying the action equ.~(\ref{s_lat_aniso}) with respect to $U_j(x)$ 
gives an equation of motion for $E_j$
\bea
E^a_j(t + \Delta t, {\vec x})  &=&  E^a_j(t, {\vec x}) 
  + \frac{i \Delta t}{g a^3} \sum_{k}  
       \left\{ {\rm Tr} \left[\lambda^a U_j(x) U_k (x+\hat{j}) 
                       U^{\dagger}_j (x +\hat{k}) U^{\dagger}_k(x) \right] 
\right. \nonumber \\ 
\label{E_eom}
&& \hspace{2cm}\left.\mbox{}
  +  {\rm Tr} \left[\lambda^a U_j(x) U^{\dagger}_{k}(x+\hat{j}-\hat{k}) 
                U^\dagger_{j}(x-\hat{k}) U_{k}( x -\hat{k}) \right] \right\}   .
\eea
We note that $E^a_j(t+\Delta t, {\vec x})$ is determined by the electric 
fields and the spatial gauge links at time $t$. Using equ.~(\ref{E_j_latt})
and the electric field $E_j^a$ at time $t+\Delta t$ we can compute the temporal 
plaquette $W_{0j}(x)$ at $t+\Delta t$. This result can be used to evolve
the spacelike gauge links
\bea
\label{U_eom}
 U_j(t+\Delta t,\vec{x})=W_{0j}(x) U_j(x)\, .
\eea
Together, equ.~(\ref{E_eom}) and equ.~(\ref{U_eom}) describe a staggered
leapfrog algorithm, similar to equ.~(\ref{leap_1}-\ref{leap_3}) above. An
important constraint on the numerical evolution is provided by Gauss law. 
Varying the lattice action with respect to $U_0$ before imposing temporal
axial gauge gives
\be
\label{Gauss_cons}
\sum_{j} \left[ E_j^a(x) -
    U^\dagger_{j}(x-\hat{j}) E_j^a(x-\hat{j}) U_j(x-\hat{j}) 
        \right] = 0 \, . 
\ee
This constraint is preserved by the evolution equations.

 The classical field equations are exactly scale invariant and there 
is no dependence on the coupling constant $g$. Physical quantities,
like the energy momentum tensor, explicitly depend on $g$. In practice
classical field simulations require a model for the initial conditions
and the corresponding coupling. The initial conditions are typically
an ensemble of gauge fields distributed according to some distribution, 
for example an anisotropic Gaussian in momentum space. The anisotropy 
is assumed to be a consequence of the strong longitudinal expansion of 
the initial state of a heavy ion collision. Physical observables are
determined by averages the evolved fields over the initial ensemble.

 Note that a purely classical field evolution does not thermalize. 
A thermal ensemble of classical fields would satisfy the equipartition
law, and the total energy would be dominated by modes near the 
lattice cutoff. This is the Rayleigh-Jeans UV catastrophe. However,
classical field evolution has interesting non-thermal fixed points
\cite{Berges:2008wm}, which may play a role in thermalization. 

 The classical field framework has been extended in a variety
of ways. One direction is the inclusion of quantum fluctuations
on top of the classical field \cite{Dusling:2010rm}. Another problem
is the inclusion of modes that are not highly populated. In the hard
thermal loop approximation one can show that hard modes can be described
as colored particles interacting with the classical field corresponding 
to the soft modes \cite{Litim:2001db}. The equations of motion for the 
colored particles are known as Wong's equations \cite{Wong:1970fu}. 
Numerical studies can be found in \cite{Hu:1996sf}.

\subsection{Nonequilibrium QCD: Holography}
\label{sec_ads}
 
 A new approach to quantum fields in and out of equilibrium is provided
by the AdS/CFT correspondence \cite{Maldacena:1997re,Son:2007vk,Gubser:2009md,CasalderreySolana:2011us,DeWolfe:2013cua}.
The AdS/CFT correspondence is a holographic duality. It asserts that the
dynamics of a quantum field theory defined on the boundary of a higher 
dimensional space is encoded in boundary correlation functions of a 
gravitational theory in the bulk. The correspondence is simplest if 
the boundary theory is strongly coupled and contains a large number 
$N$ of degrees of freedom. In this case the bulk theory is simply classical 
Einstein gravity. The partition function of the boundary quantum 
field theory (QFT) is 
\be 
 Z_{\it QFT}[J_i]=\exp\left(-S\left[\left.\phi_i\right|_{\partial{\it M}}
= J_i\right]\right)\, , 
\ee
where $J_i$ is a set of sources in the field theory, $S$ is the gravitational 
action, $\phi_i$ is a dual set of fields in the gravitational theory, and 
$\partial{\it M}$ is the boundary of $AdS_5$. The fields $\phi_i$ satisfy 
classical equations of motions subject to boundary conditions on 
$\partial{\it M}$.

 The original construction involves a black hole in AdS$_5$ and is dual 
to a relativistic fluid governed by a generalization of QCD known as 
${\cal N}=4$ super Yang-Mills theory. This theory is considered in 
the limit of a large number of colors $N_c$. The gravitational theory 
is Einstein gravity with additional matter fields that are not 
relevant here. The AdS$_5$ black hole metric is  
\be
\label{bh_son}
ds^2 = \frac{(\pi T R_a)^2}{u}  \left(-f(u) dt^2 + d\vec{x}^2 \right) + 
  \frac{R_a^2}{4 u^2 f(u)} du^2\, ,
\ee
where $\vec{x},t$ are Minkowski space coordinates, and $u$ is a 
``radial'' coordinate where $u=1$ is the location of the black hole 
horizon and $u=0$ is the boundary. $T$ is the temperature, $R_a$ is the 
AdS radius, and $f(u)=1-u^2$. 

 It is instructive to check that this metric does indeed provide a 
solution to the Einstein equations with a negative cosmological 
constant. This can be done using a simple Mathematica script. I begin
by defining the metric and its inverse: 

\vspace*{0.3cm} 
\begin{lstlisting}
(* metric *)
(* ------ *)
n = 5;
coord = {t, x, y, z, u};
f[u_] := 1 - u^2
metric = DiagonalMatrix[{-f[u]/u*(Pi*T*Ra)^2, (Pi*T*Ra)^2/u, (Pi*T*Ra)^2/
    u, (Pi*T*Ra)^2/u, Ra^2/(4*u^2*f[u])}]
inversemetric = Simplify[Inverse[metric]]
\end{lstlisting}

\vspace*{0.3cm} 
\noindent
From the metric I compute the Christoffel symbols 
\be
\Gamma^\mu_{\alpha\beta} = \frac{1}{2}g^{\mu\nu}
\left( \partial_\alpha g_{\nu\beta}
      +\partial_\beta  g_{\nu\alpha}
      -\partial_\mu    g_{\alpha\beta}\right) \, , 
\ee
the Riemann tensor 
\be 
R^\mu_{\;\, \nu\alpha\beta} = 
  \partial_\alpha \Gamma^\mu_{\nu\beta} 
- \partial_\beta  \Gamma^\mu_{\nu\alpha}
+ \Gamma^\rho_{\nu\beta}\Gamma^\mu_{\rho\alpha}
- \Gamma^\rho_{\nu\alpha}\Gamma^\mu_{\rho\beta} \, , 
\ee
the Ricci tensor $R_{\alpha\beta}=R^\mu_{\;\,\alpha\mu\beta}$, and the 
scalar curvature $R=R^\mu_{\;\,\mu}$. Finally, I compute the 
Einstein tensor $G_{\mu\nu}=R_{\mu\nu}-\frac{1}{2}g_{\mu\nu}R$. 

\vspace*{0.3cm} 
\begin{lstlisting}
(* Christoffel Symbols *)
(* ------------------- *)
affine :=  affine = Simplify[
   Table[(1/2)*
     Sum[(inversemetric[[i, s]])*(D[metric[[s, j]], coord[[k]]] + 
         D[metric[[s, k]], coord[[j]]] - 
         D[metric[[j, k]], coord[[s]]]), {s, 1, n}], {i, 1, n}, {j, 1,
      n}, {k, 1, n}]]

(* Riemann Tensor *)
(* -------------- *)
riemann :=  riemann =   Simplify[Table[
    D[affine[[i, j, l]], coord[[k]]] - 
     D[affine[[i, j, k]], coord[[l]]] + 
     Sum[affine[[s, j, l]]*affine[[i, k, s]] - 
       affine[[s, j, k]]*affine[[i, l, s]], {s, 1, n}], {i, 1, n}, {j,
      1, n}, {k, 1, n}, {l, 1, n}]]

(* Ricci Tensor *)
(* ------------ *)
ricci :=  ricci = Simplify[
   Table[Sum[riemann[[i, j, i, l]], {i, 1, n}], {j, 1, n}, {l, 1, n}]]

(* scalar curvature *)
(* ---------------- *)
scalar = Simplify[
  Sum[inversemetric[[i, j]]*ricci[[i, j]], {i, 1, n}, {j, 1, n}]]

(* Einstein tensor *)
(* --------------- *)
einstein = Simplify[ricci - (1/2)*scalar*metric]
\end{lstlisting}

\vspace*{0.3cm} 
Now I can check the equation of motion, $G_{\mu\nu}=\frac{\Lambda}{2}
g_{\mu\nu}$, where the cosmological constant is determined by the 
AdS radius $R$. 

\begin{lstlisting}
(* Field equation with cosmological constant *)
(* ----------------------------------------- *)
lam = 12/Ra^2;
Simplify[einstein - lam/2*metric]
\end{lstlisting}

\vspace*{0.3cm} 
 In the boundary theory the metric couples to the stress tensor 
$\Pi_{\mu\nu}$. Correlation functions of the stress tensor can be found 
by linearizing the bulk action around the AdS$_5$ solution, $g_{\mu\nu}=
g_{\mu\nu}^0+\delta g_{\mu\nu}$. Small oscillations of the off-diagonal 
strain $\delta g_x^y$ are particularly simple, because the equation of 
motion for $\phi\equiv g_x^y$ is that of a minimally coupled scalar
\be 
\frac{1}{\sqrt{-g}}\partial_\mu \left( \sqrt{-g}g^{\mu\nu}\partial_\nu 
 \phi\right) = 0\, .  
\ee 
The wave equation can be obtained using the metric coefficients defined
above. 

\vspace*{0.3cm} 
\begin{lstlisting}
(* \sqrt{-g} g^{\mu\nu} \partial_{nu} \Phi(t,z,u) *)
(* -------------------------------------------- *)
SqrtG = Simplify[Sqrt[-Det[metric]], {Ra > 0, T > 0, u > 0}]
dnuPhi = Table[D[Phi[t, z, u], coord[[i]]], {i, 1, n}];
DnuPhi = SqrtG*inversemetric.dnuPhi;

(* Laplacian, up to factor \sqrt{-g} *)
(* --------------------------------- *)
DPhi = FullSimplify[Sum[D[DnuPhi[[nu]], coord[[nu]]], {nu, 1, n}]]

(* harmonic space and time dependence *)
(* ---------------------------------- *)
DPhiS = 
 DPhi /. { D[Phi[t, z, u], {z, 2}] -> -k^2*fp, 
   D[Phi[t, z, u], {t, 2}] -> -w^2*fp, 
   D[Phi[t, z, u], {u, 2}] -> fpPP, D[Phi[t, z, u], {u, 1}] -> fpP}
\end{lstlisting}

\vspace*{0.3cm} 
In the case of harmonic dependence on the Minkowski coordinates $\delta g_x^y
=\phi_k(u)e^{ikx-i\omega t}$ the fluctuations are governed by the wave equation
\be
\label{lingrav}
\phi_k''(u) - \frac{1+u^2}{uf(u)} \phi_k'(u) 
      + \frac{\omega^2 -k^2f(u)}{(2\pi T)^2 u f(u)^2}
  \phi_k(u) = 0\, .
\ee
This differential equation has two linearly independent solutions.
The retarded correlation function corresponds to picking a solution 
that is purely infalling at the horizon \cite{Son:2007vk}. The retarded 
correlation function $G_R(\omega,k)$ defined in equ.~(\ref{G_ret})
is determined by inserting the solution into the Einstein-Hilbert action, 
and then computing the variation with respect to the boundary value 
of $\delta g_x^y$. 

\begin{figure}[t!]
\begin{center}
\includegraphics*[width=6.5cm]{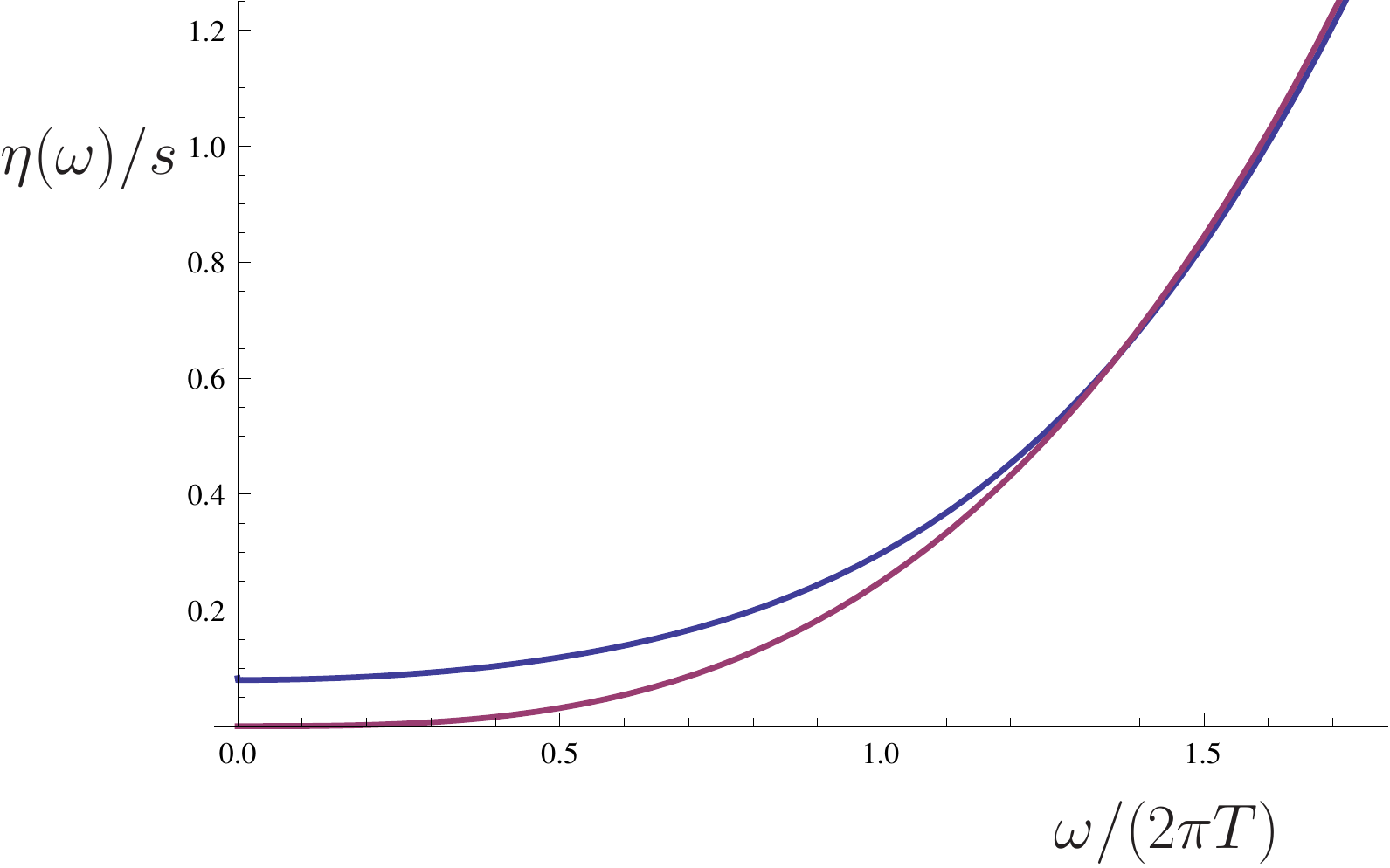}
\includegraphics*[width=6.5cm]{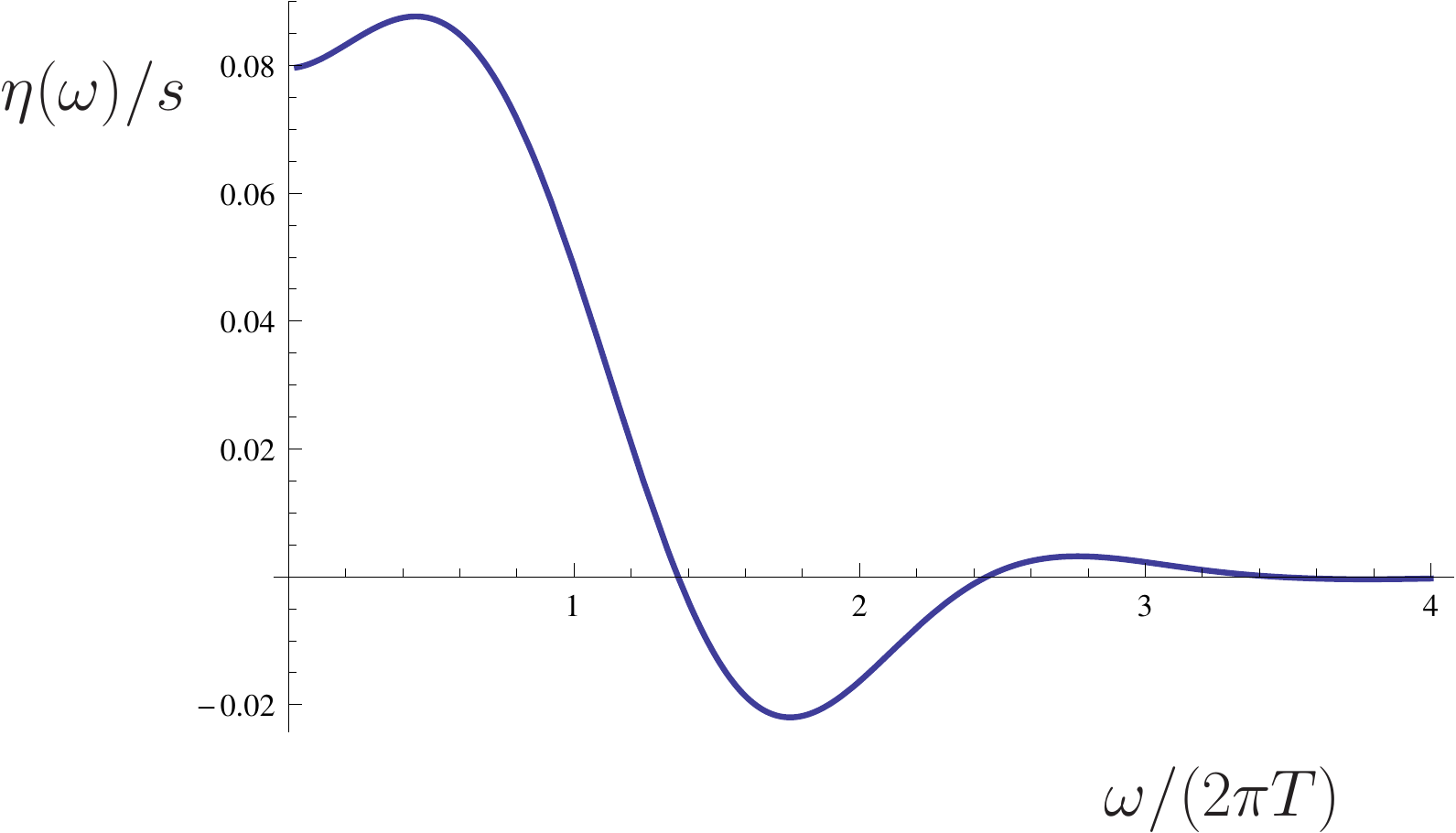}
\end{center}
\caption{\label{fig_ads_spec_fct}
Viscosity spectral function in a ${\cal N}=4$ SUSY Yang Mills plasma. 
The spectral function is computed in the large $N_c$ limit of a strongly 
coupled plasma using the AdS/CFT correspondence. The figure in the left 
panel shows $\eta(\omega)/s$ (blue) and the zero temperature counterpart 
$\eta_{T=0}(\omega)/s$ (red) as a function of $\omega$. The figure in the 
right panel shows the finite temperature part $[\eta(\omega)-\eta_{T=0}
(\omega)]/s$. The figures were generated using the script described 
below equ.~(\ref{G_R_Action}).}
\end{figure}

 The infalling solution can be expressed as 
\be 
\label{phi_k_ans}
\phi_k(u) = (1-u) ^{-i\mathfrak{w}/2}F_k(u)
\ee
where $\mathfrak{w}=\omega/(2\pi T)$ and the first factor describes
the near horizon behavior. The function $F_k(u)$ can be obtained
as an expansion in $\mathfrak{w}$ and $\mathfrak{k}=k/(2\pi T)$. 
At second order in $O(\mathfrak{w}$ and $\mathfrak{k}$ the solution 
is \cite{Policastro:2002se}
\be 
F_k(u) = 1-\frac{i\mathfrak{w}}{2}\log\left(\frac{1+u}{2}\right)
  +\frac{\mathfrak{w}^2}{8} \left\{
     \left[ 8 - \frac{8\mathfrak{k}^2}{\mathfrak{w}^2}
              + \log\left(\frac{1+u}{2}\right)\right]
                              \log\left(\frac{1+u}{2}\right)
   - 4{\it Li}_2\left(\frac{1-u}{2}\right)\right\}\, . 
\ee
In the opposite limit, $\mathfrak{w}\gg 1$, the wave equation can be 
solved using a WKB approximation \cite{Teaney:2006nc}. For $\mathfrak{k}=0$ 
the result is 
\be 
\phi_k(u) = \pi\mathfrak{w}^2\frac{u}{\sqrt{1-u^2}}
 \left[ iJ_2\left(2\mathfrak{w}\sqrt{u}\right)
         -Y_2\left(2\mathfrak{w}\sqrt{u}\right)\right]\, . 
\ee
In the intermediate regime the wave equation can be solved numerically.
A standard method is to start from the near horizon result given in 
equ.~(\ref{phi_k_ans}) and integrate outwards towards the boundary. The 
retarded correlation function is given by the variation of the boundary 
action with respect to the field. For this purpose we consider the 
quadratic part of the Einstein-Hilbert action and use the AdS/CFT 
correspondence to express Newton's constant in terms of gauge theory 
parameters. We find
\be 
 S = -\frac{\pi^2N^2T^4}{8}\int du\int d^4x\, 
    \frac{f(u)}{u}  \left(\partial_u\phi\right)^2 + \ldots \, .  
\ee
The boundary action follows after an integration by parts. The retarded 
Green function is determined by the second variational derivative with 
respect to the boundary value of the field 
\cite{Policastro:2002se,Son:2006em}, 
\be 
\label{G_R_Action}
G_R(\mathfrak{w},\mathfrak{k})= -\frac{\pi^2N^2T^4}{4}
  \left[ \frac{f(u)\partial_u \phi_k(u)}{u\phi_k(u)}
     \right]_{u\to 0}\, .
\ee
Finally, the spectral function is given by $\eta(\omega,k)=-\omega^{-1}
{\it Im}\,G_R(\omega,k)$. Below is a short Mathematica script that 
determines the spectral function numerically.

\vspace*{0.3cm} 
\begin{lstlisting}
(* equation of motion for minimally coupled scalar *)
(* with harmonic space and time dependence         *)
(* ----------------------------------------------- *)
f[u_] := 1 - u^2
EomPhi = phi''[u] - (1 + u^2)/(u f[u]) phi'[u] 
   + (w^2 - q^2 f[u])/(u f[u]^2) phi[u]

(* boundary solution *)
(* ----------------- *)
phiHorizon[u_] := (1-u)^(-I*w/2)

(* numerically integrate from Horizon to boundary *)
(* ---------------------------------------------- *)
SolPhi[omega_, qq_] := Block[{w = omega, q = qq},
   NDSolve[
    {0 == EomPhi,
     phi[epsH]  == phiHorizon[epsH],
     phi'[epsH] == phiHorizon'[epsH]},
    phi[u],
    {u, epsB, epsH}]][[1, 1, 2]]

(* retarded correlator from boundary action *)
(* ---------------------------------------- *)
Gret[omega_, qq_] := (f[u]/u D[solPhi[omega, qq], u]/solPhi[omega, qq] ) 
 /. {u -> epsB}

\end{lstlisting}

\begin{figure}[t!]
\begin{center}
\includegraphics*[width=8.5cm]{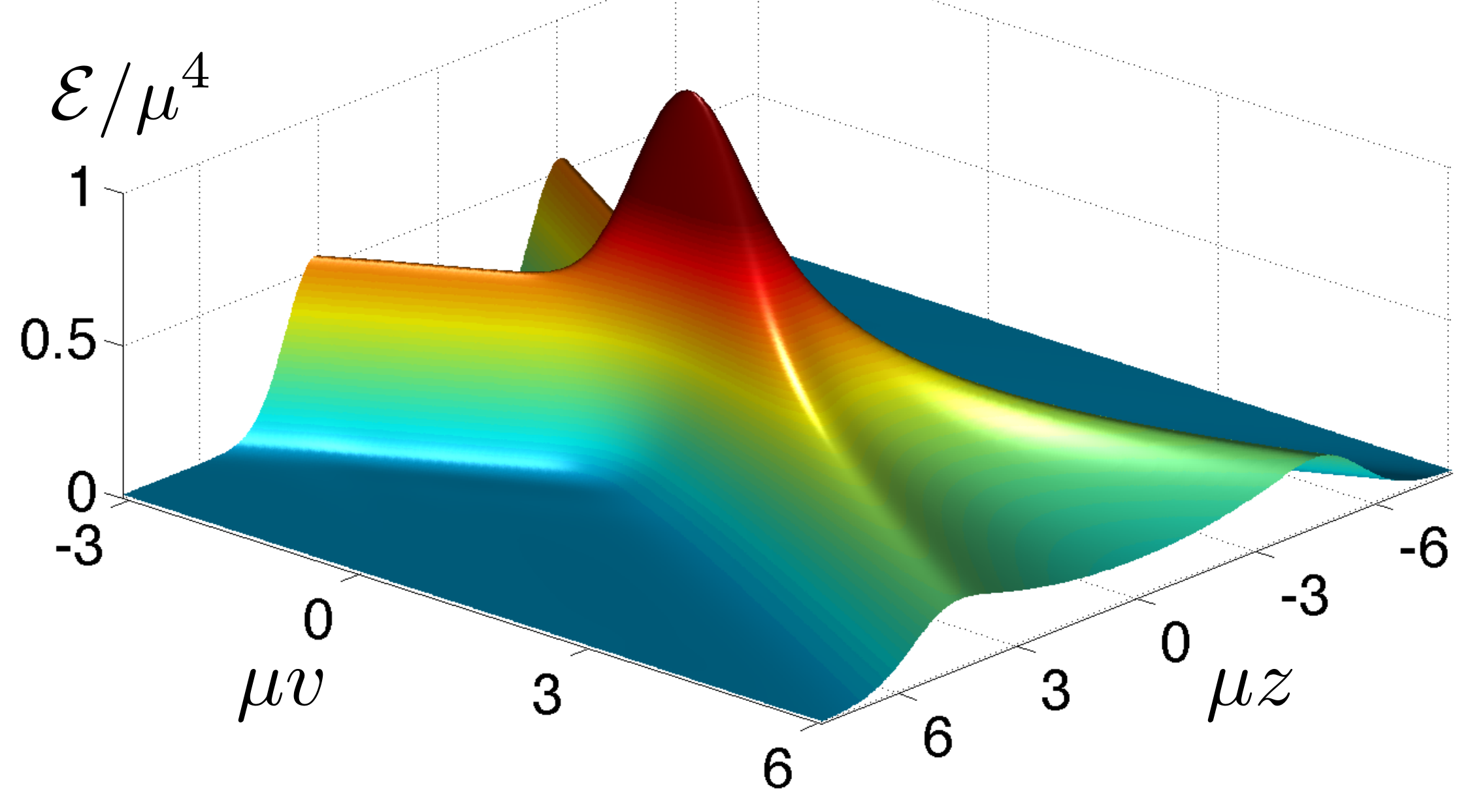}
\end{center}
\caption{\label{fig_coll_shocks}
Energy density of colliding shock waves in $AdS_5$ space \cite{Chesler:2010bi}.
The figure shows the energy density ${\cal E}/\mu^4$ on the boundary of 
$AdS_5$ as a function of the time coordinate $v$ and the longitudinal 
direction $z$. The shocks are infinitely extended in the transverse 
direction. The parameter $\mu$ sets the overall scale. }
\end{figure}

\vspace*{0.3cm} 
The spectral function for $k=0$ is shown in Fig.~\ref{fig_ads_spec_fct}. This 
is an interesting result because it represent a systematic calculation of a
real time observable in the strong coupling limit of a quantum field theory. 
As explained in Sect.~\ref{sec_lQCD_real} the corresponding lattice 
calculation is very difficult, and existing results are difficult to improve 
upon. We also note that the result is quite different from expectations at 
weak coupling. At weak coupling we expect the spectral function to show a 
narrow transport peak at zero energy \cite{Schaefer:2014awa}.

 So far we have only considered calculations very close to equilibrium, 
corresponding to small perturbations of the $AdS_5$ Schwarzschild solution. 
In order to address the problem of initial state dynamics and thermalization
we have to consider initial conditions that mimic colliding nuclei. Recent
work focuses on colliding shock waves in asymptotically $AdS_5$ spaces. 
In the strong coupling limit the evolution of the shock waves is a problem
in numerical relativity. Special methods have been developed to deal 
with problems in $AdS$ space \cite{Chesler:2013lia}. These methods are 
quite different from the techniques employed in connection with black
hole or neutron star mergers in asymptotically flat Minkowski space time. 
A typical result is shown in Fig.~\ref{fig_coll_shocks}. The calculations
demonstrate fast ``hydrodynamization'', that means a rapid decay of 
non-hydrodynamic modes. At somewhat longer time scales thermal 
equilibration is achieved. This corresponds to the formation of an 
event horizon in the bulk. In general, it was realized that there 
is a fluid-gravity correspondence, an equivalence between dynamic 
space times containing a horizon and solutions of the Navier-Stokes
equation \cite{Rangamani:2009xk}. This correspondence can be used 
to study, both analytically and numerically, difficult problems in 
fluid dynamics.  

\section{Outlook and acknowledgments}
\label{sec_out}

 I hope this brief review provides a flavor of the breadth of computational
problems that are related QCD. This includes many issues that are at the 
forefront of computational physics, like the sign problem in euclidean 
QCD at finite baryon density, and the challenge to extract real time 
correlation functions from the euclidean path integral. It also includes
many problems that are of great interest to mathematicians. Both the 
Yang-Mills existence and mass gap as well as the Navier-Stokes existence 
and smoothness problems are among the Clay Millenium Prize problems
\cite{Clay-YM,Clay-NS}. Interesting work on the Boltzmann equation was 
recently recognized with a Fields medal \cite{Villani:2009}, and gradient 
flow plays an important role in the proof of the Poincare conjecture
\cite{Perelman}.

\begin{acknowledgement}
The euclidean path integral simulation in quantum
mechanics is described in \cite{Schafer:2004xa}, and the programs are 
available at {\tt https://www.physics.ncsu.edu/schaefer/{\allowbreak}physics/}. 
A simple $Z_2$ lattice gauge code can be found in the Appendix. You should be
able to extend this code to $SU(2)$ and $SU(3)$. Modern lattice QCD tools
can be found on the chroma website {\tt http://github.com/JeffersonLab/chroma}.
The VH1 hydro code is described in \cite{Blondin:1993} and can be downloaded
at {\tt http://wonka.{\allowbreak}physics.ncsu.edu/pub/VH-1/}. Dissipative 
and anisotropic versions are available on request. There are a number of 
relativistic hydro codes on the web. An example is the VISHNU code 
\cite{Shen:2014vra} which is available at {\tt https://u.osu.edu/vishnu/}. 
Both UrQMD {\tt http://urqmd.org/} and GiBUU 
{\tt https://gibuu.{\allowbreak}hepforge.org/} are also available online.

 The mathematica notebooks in Sect.~\ref{sec_ads} are adapted from
notebooks available on Jim Hartle's website 
{\tt http://web.{\allowbreak}physics.ucsb.edu/{\textasciitilde}gravitybook/}. 
Much more sophisticated tensor packages are easily found on the web. 
The simple script for solving the wave equation in $AdS_5$ is adapted 
from a notebook written by Matthias Kaminski. A set of lecture notes 
and mathematica notebooks for solving the Einstein equations numerically
on asymptotically $AdS$ spaces can be found on Wilke van der Schee's website
{\tt https://sites.google.com/site/{\allowbreak}wilkevanderschee/ads-numerics}.
T.~S. work is supported by the US Department 
of Energy grant DE-FG02-03ER41260. 

\end{acknowledgement}

\section*{Appendix: $Z_2$ gauge theory}
\addcontentsline{toc}{section}{Appendix}
 This is a simple Monte Carlo program for $Z_2$ gauge theory 
written by M.~Creutz \cite{Creutz:2004}.

\begin{lstlisting}
/* Z_2 lattice gauge simulation */
/* Michael Creutz <creutz@bnl.gov>     */
/* http://thy.phy.bnl.gov/~creutz/z2.c */

#include <stdio.h>
#include <stdlib.h>
#include <math.h>

/* the lattice is of dimensions SIZE**4  */
#define SIZE 6
int link[SIZE][SIZE][SIZE][SIZE][4]; /* last index gives link direction */

/* utility functions */
void moveup(int x[],int d) {
  x[d]+=1;
  if (x[d]>=SIZE) x[d]-=SIZE; 
  return;
}
void movedown(int x[],int d) {
  x[d]-=1;
  if (x[d]<0) x[d]+=SIZE;
  return;
}
void coldstart(){  /* set all links to unity */
  int x[4],d;
  for (x[0]=0;x[0]<SIZE;x[0]++)
    for (x[1]=0;x[1]<SIZE;x[1]++)
      for (x[2]=0;x[2]<SIZE;x[2]++)
        for (x[3]=0;x[3]<SIZE;x[3]++)
          for (d=0;d<4;d++)
	    link[x[0]][x[1]][x[2]][x[3]][d]=1;
  return;
}
/* for a random start: call coldstart() and then update once at beta=0 */

/* do a Monte Carlo sweep; return energy */
double update(double beta){
  int x[4],d,dperp,staple,staplesum;    
  double bplus,bminus,action=0.0; 
  for (x[0]=0; x[0]<SIZE; x[0]++)
    for (x[1]=0; x[1]<SIZE; x[1]++)
      for (x[2]=0; x[2]<SIZE; x[2]++)
        for (x[3]=0; x[3]<SIZE; x[3]++)
          for (d=0; d<4; d++) {
            staplesum=0;
            for (dperp=0;dperp<4;dperp++){
              if (dperp!=d){
                /*  move around thusly:
                    dperp        6--5
                    ^            |  |
                    |            1--4
                    |            |  |
                    -----> d     2--3  */
                /* plaquette 1234 */
                movedown(x,dperp);
                staple=link[x[0]][x[1]][x[2]][x[3]][dperp]
                  *link[x[0]][x[1]][x[2]][x[3]][d];
                moveup(x,d);
                staple*=link[x[0]][x[1]][x[2]][x[3]][dperp];  
                moveup(x,dperp);
                staplesum+=staple;
                /* plaquette 1456 */
                staple=link[x[0]][x[1]][x[2]][x[3]][dperp];
                moveup(x,dperp);
                movedown(x,d);
                staple*=link[x[0]][x[1]][x[2]][x[3]][d];
                movedown(x,dperp);
                staple*=link[x[0]][x[1]][x[2]][x[3]][dperp];
                staplesum+=staple;
              }
	    }
            /* calculate the Boltzmann weight */
            bplus=exp(beta*staplesum);
            bminus=1/bplus;
            bplus=bplus/(bplus+bminus);
            /* the heatbath algorithm */
            if ( drand48() < bplus ){
              link[x[0]][x[1]][x[2]][x[3]][d]=1;
              action+=staplesum;
            }
            else{ 
              link[x[0]][x[1]][x[2]][x[3]][d]=-1;
              action-=staplesum;
            }
          }
  action /= (SIZE*SIZE*SIZE*SIZE*4*6);
  return 1.-action;
}

/******************************/
int main(){
  double beta, dbeta, action;
  srand48(1234L);  /* initialize random number generator */
  /* do your experiment here; this example is a thermal cycle */
  dbeta=.01;
  coldstart();
  /* heat it up */
  for (beta=1; beta>0.0; beta-=dbeta){
    action=update(beta);
    printf("%g\t%g\n",beta,action); 
  }
  printf("\n\n");
  /* cool it down */
  for (beta=0; beta<1.0; beta+=dbeta){
    action=update(beta);
    printf("%g\t%g\n",beta,action); 
  }
  printf("\n\n");
  exit(0);
}

\end{lstlisting}
\newpage
\bibliographystyle{spphys}
\bibliography{lib-Thomas}

\begin{thebibliography}{100}
\providecommand{\url}[1]{{#1}}
\providecommand{\urlprefix}{URL }
\expandafter\ifx\csname urlstyle\endcsname\relax
  \providecommand{\doi}[1]{DOI \discretionary{}{}{}#1}\else
  \providecommand{\doi}{DOI \discretionary{}{}{}\begingroup
  \urlstyle{rm}\Url}\fi

\bibitem{Feynman}
R.P. Feynman, A.R. Hibbs, \emph{Quantum Mechanics and Path Integrals}
  (McGraw-Hill, 1965)

\bibitem{Metropolis:1953am}
N.~Metropolis, A.W. Rosenbluth, M.N. Rosenbluth, A.H. Teller, E.~Teller, J.
  Chem. Phys. \textbf{21}, 1087 (1953).
\newblock \doi{10.1063/1.1699114}

\bibitem{Creutz:1980gp}
M.~Creutz, B.~Freedman, Annals Phys. \textbf{132}, 427 (1981).
\newblock \doi{10.1016/0003-4916(81)90074-9}

\bibitem{Shuryak:1984xr}
E.V. Shuryak, O.V. Zhirov, Nucl. Phys. \textbf{B242}, 393 (1984).
\newblock \doi{10.1016/0550-3213(84)90401-2}

\bibitem{Shuryak:1987tr}
E.V. Shuryak, Nucl. Phys. \textbf{B302}, 621 (1988).
\newblock \doi{10.1016/0550-3213(88)90191-5}

\bibitem{Schafer:2004xa}
T.~Sch{\"a}fer, {Instantons and Monte Carlo methods in quantum mechanics}
  (2004).
\newblock ArXive:hep-lat/0411010

\bibitem{Jarrell:1996rrw}
M.~Jarrell, J.E. Gubernatis, Phys. Rept. \textbf{269}, 133 (1996).
\newblock \doi{10.1016/0370-1573(95)00074-7}

\bibitem{Asakawa:2000tr}
M.~Asakawa, T.~Hatsuda, Y.~Nakahara, Prog. Part. Nucl. Phys. \textbf{46}, 459
  (2001).
\newblock \doi{10.1016/S0146-6410(01)00150-8}

\bibitem{Jarzynski:1997}
C.~{Jarzynski}, Physical Review Letters \textbf{78}, 2690 (1997).
\newblock \doi{10.1103/PhysRevLett.78.2690}

\bibitem{Gross:1973id}
D.J. Gross, F.~Wilczek, Phys.Rev.Lett. \textbf{30}, 1343 (1973).
\newblock \doi{10.1103/PhysRevLett.30.1343}

\bibitem{Politzer:1973fx}
H.D. Politzer, Phys.Rev.Lett. \textbf{30}, 1346 (1973).
\newblock \doi{10.1103/PhysRevLett.30.1346}

\bibitem{Coleman:1973jx}
S.R. Coleman, E.J. Weinberg, Phys.Rev. \textbf{D7}, 1888 (1973).
\newblock \doi{10.1103/PhysRevD.7.1888}

\bibitem{Nakamura:2010zzi}
K.~Nakamura, et~al., J. Phys. \textbf{G37}, 075021 (2010).
\newblock \doi{10.1088/0954-3899/37/7A/075021}

\bibitem{Alford:2007xm}
M.G. Alford, A.~Schmitt, K.~Rajagopal, T.~Sch{\"a}fer, Rev. Mod. Phys.
  \textbf{80}, 1455 (2008).
\newblock \doi{10.1103/RevModPhys.80.1455}

\bibitem{Adams:2012th}
A.~Adams, L.D. Carr, T.~Sch{\"a}fer, P.~Steinberg, J.E. Thomas, New J. Phys.
  \textbf{14}, 115009 (2012).
\newblock \doi{10.1088/1367-2630/14/11/115009}

\bibitem{Braun-Munzinger:2015hba}
P.~Braun-Munzinger, V.~Koch, T.~Sch{\"a}fer, J.~Stachel, Phys. Rept.
  \textbf{621}, 76 (2016).
\newblock \doi{10.1016/j.physrep.2015.12.003}

\bibitem{GellMann:1968rz}
M.~Gell-Mann, R.J. Oakes, B.~Renner, Phys. Rev. \textbf{175}, 2195 (1968).
\newblock \doi{10.1103/PhysRev.175.2195}

\bibitem{Coleman:1980mx}
S.R. Coleman, E.~Witten, Phys. Rev. Lett. \textbf{45}, 100 (1980).
\newblock \doi{10.1103/PhysRevLett.45.100}

\bibitem{tHooft:1979bh}
G.~'t~Hooft, NATO Sci. Ser. B \textbf{59}, 135 (1980)

\bibitem{Shuryak:1977ut}
E.V. Shuryak, Sov. Phys. JETP \textbf{47}, 212 (1978).
\newblock [Zh. Eksp. Teor. Fiz.74,408(1978)]

\bibitem{Shuryak:1978ij}
E.V. Shuryak, Phys. Lett. \textbf{B78}, 150 (1978).
\newblock \doi{10.1016/0370-2693(78)90370-2}.
\newblock [Yad. Fiz.28,796(1978)]

\bibitem{Linde:1980ts}
A.D. Linde, Phys. Lett. \textbf{B96}, 289 (1980).
\newblock \doi{10.1016/0370-2693(80)90769-8}

\bibitem{Pisarski:1983ms}
R.D. Pisarski, F.~Wilczek, Phys. Rev. \textbf{D29}, 338 (1984).
\newblock \doi{10.1103/PhysRevD.29.338}

\bibitem{Aoki:2006we}
Y.~Aoki, G.~Endrodi, Z.~Fodor, S.D. Katz, K.K. Szabo, Nature \textbf{443}, 675
  (2006).
\newblock \doi{10.1038/nature05120}

\bibitem{Bazavov:2011nk}
A.~Bazavov, et~al., Phys. Rev. \textbf{D85}, 054503 (2012).
\newblock \doi{10.1103/PhysRevD.85.054503}

\bibitem{Aoki:2006br}
Y.~Aoki, Z.~Fodor, S.D. Katz, K.K. Szabo, Phys. Lett. \textbf{B643}, 46 (2006).
\newblock \doi{10.1016/j.physletb.2006.10.021}

\bibitem{Aoki:2009sc}
Y.~Aoki, S.~Borsanyi, S.~Durr, Z.~Fodor, S.D. Katz, S.~Krieg, K.K. Szabo, JHEP
  \textbf{06}, 088 (2009).
\newblock \doi{10.1088/1126-6708/2009/06/088}

\bibitem{Bazavov:2014pvz}
A.~Bazavov, et~al., Phys. Rev. \textbf{D90}(9), 094503 (2014).
\newblock \doi{10.1103/PhysRevD.90.094503}

\bibitem{Stephanov:2004wx}
M.A. Stephanov, Prog. Theor. Phys. Suppl. \textbf{153}, 139 (2004)

\bibitem{Fodor:2001pe}
Z.~Fodor, S.D. Katz, JHEP \textbf{03}, 014 (2002)

\bibitem{Allton:2002zi}
C.~Allton, S.~Ejiri, S.~Hands, O.~Kaczmarek, F.~Karsch, et~al., Phys.Rev.
  \textbf{D66}, 074507 (2002).
\newblock \doi{10.1103/PhysRevD.66.074507}

\bibitem{Karsch:2003va}
F.~Karsch, C.R. Allton, S.~Ejiri, S.J. Hands, O.~Kaczmarek, E.~Laermann,
  C.~Schmidt, Nucl. Phys. Proc. Suppl. \textbf{129}, 614 (2004).
\newblock \doi{10.1016/S0920-5632(03)02659-8}.
\newblock [,614(2003)]

\bibitem{Fodor:2004nz}
Z.~Fodor, S.~Katz, JHEP \textbf{0404}, 050 (2004).
\newblock \doi{10.1088/1126-6708/2004/04/050}

\bibitem{Gavai:2008zr}
R.V. Gavai, S.~Gupta, Phys. Rev. \textbf{D78}, 114503 (2008).
\newblock \doi{10.1103/PhysRevD.78.114503}

\bibitem{Datta:2012pj}
S.~Datta, R.V. Gavai, S.~Gupta, Nucl. Phys. \textbf{A904-905}, 883c (2013).
\newblock \doi{10.1016/j.nuclphysa.2013.02.156}

\bibitem{deForcrand:2010he}
P.~de~Forcrand, O.~Philipsen, Phys. Rev. Lett. \textbf{105}, 152001 (2010).
\newblock \doi{10.1103/PhysRevLett.105.152001}

\bibitem{Stephanov:1998dy}
M.A. Stephanov, K.~Rajagopal, E.V. Shuryak, Phys. Rev. Lett. \textbf{81}, 4816
  (1998)

\bibitem{Sauer:1976zzf}
G.~Sauer, H.~Chandra, U.~Mosel, Nucl. Phys. \textbf{A264}, 221 (1976).
\newblock \doi{10.1016/0375-9474(76)90429-2}

\bibitem{Pochodzalla:1995xy}
J.~Pochodzalla, et~al., Phys. Rev. Lett. \textbf{75}, 1040 (1995).
\newblock \doi{10.1103/PhysRevLett.75.1040}

\bibitem{Elliott:2013pna}
J.B. Elliott, P.T. Lake, L.G. Moretto, L.~Phair, Phys. Rev. \textbf{C87}(5),
  054622 (2013).
\newblock \doi{10.1103/PhysRevC.87.054622}

\bibitem{Alford:1998mk}
M.G. Alford, K.~Rajagopal, F.~Wilczek, Nucl. Phys. \textbf{B537}, 443 (1999).
\newblock \doi{10.1016/S0550-3213(98)00668-3}

\bibitem{Schafer:1999fe}
T.~Sch{\"a}fer, Nucl. Phys. \textbf{B575}, 269 (2000).
\newblock \doi{10.1016/S0550-3213(00)00063-8}

\bibitem{Schafer:1998ef}
T.~Sch{\"a}fer, F.~Wilczek, Phys. Rev. Lett. \textbf{82}, 3956 (1999).
\newblock \doi{10.1103/PhysRevLett.82.3956}

\bibitem{Hatsuda:2006ps}
T.~Hatsuda, M.~Tachibana, N.~Yamamoto, G.~Baym, Phys. Rev. Lett. \textbf{97},
  122001 (2006).
\newblock \doi{10.1103/PhysRevLett.97.122001}

\bibitem{Creutz:1983}
M.~Creutz, \emph{Quarks, Gluons, and Lattices} (Cambridge University Press,
  1983)

\bibitem{Montvay:1994}
I.~Montvay, G.~M{\"u}nster, \emph{Quantum Fields on a Lattice} (Cambridge
  University Press, 1994)

\bibitem{Smit:2002}
J.~Smit, \emph{Introduction to Quantum Fields on a Lattice} (Cambridge
  University Press, 2002)

\bibitem{Gattringer:2009}
C.~Gattringer, C.B. Lang, \emph{Quantum Chromodynamics on the Lattice}
  (Springer, 2009)

\bibitem{Lin:2014}
H.W. Lin, H.B. Meyer, \emph{Lattice QCD for Nuclear Physics} (Springer, 2014)

\bibitem{Fodor:2009ax}
Z.~Fodor, S.D. Katz, {The Phase diagram of quantum chromodynamics} (2009).
\newblock ArXiv:0908.3341

\bibitem{Ding:2015ona}
H.T. Ding, F.~Karsch, S.~Mukherjee, {Thermodynamics of strong-interaction
  matter from Lattice QCD} (2015).
\newblock ArXiv:1504.05274

\bibitem{Wilson:1974sk}
K.G. Wilson, Phys. Rev. \textbf{D10}, 2445 (1974).
\newblock \doi{10.1103/PhysRevD.10.2445}.
\newblock [,45(1974)]

\bibitem{Mezzadri:2006}
F.~{Mezzadri}, {How to generate random matrices from the classical compact
  groups} (2006).
\newblock ArXiv:math-ph/0609050

\bibitem{Hasenfratz:1980kn}
A.~Hasenfratz, P.~Hasenfratz, Phys. Lett. \textbf{B93}, 165 (1980).
\newblock \doi{10.1016/0370-2693(80)90118-5}.
\newblock [,241(1980)]

\bibitem{Lepage:1998dt}
G.P. Lepage, in \emph{{Strong interactions at low and intermediate energies.
  Proceedings, 13th Annual Hampton University Graduate Studies, HUGS'98,
  Newport News, USA, May 26-June 12, 1998}} (1998), pp. 49--90

\bibitem{Wolff:1988uh}
U.~Wolff, Phys. Rev. Lett. \textbf{62}, 361 (1989).
\newblock \doi{10.1103/PhysRevLett.62.361}

\bibitem{Kogut:1974ag}
J.B. Kogut, L.~Susskind, Phys. Rev. \textbf{D11}, 395 (1975).
\newblock \doi{10.1103/PhysRevD.11.395}

\bibitem{Kaplan:1992bt}
D.B. Kaplan, Phys. Lett. \textbf{B288}, 342 (1992).
\newblock \doi{10.1016/0370-2693(92)91112-M}

\bibitem{Neuberger:1997fp}
H.~Neuberger, Phys. Lett. \textbf{B417}, 141 (1998).
\newblock \doi{10.1016/S0370-2693(97)01368-3}

\bibitem{Luscher:2010ae}
M.~Luscher, in \emph{{Modern perspectives in lattice QCD: Quantum field theory
  and high performance computing. Proceedings, International School, 93rd
  Session, Les Houches, France, August 3-28, 2009}} (2010), pp. 331--399

\bibitem{Endress:2014qpa}
E.~Endress, C.~Pena, K.~Sivalingam, Comput. Phys. Commun. \textbf{195}, 35
  (2015).
\newblock \doi{10.1016/j.cpc.2015.04.017}

\bibitem{Dick:2015twa}
V.~Dick, F.~Karsch, E.~Laermann, S.~Mukherjee, S.~Sharma, Phys. Rev.
  \textbf{D91}(9), 094504 (2015).
\newblock \doi{10.1103/PhysRevD.91.094504}

\bibitem{Teper:1985ek}
M.~Teper, Phys. Lett. \textbf{B171}, 86 (1986).
\newblock \doi{10.1016/0370-2693(86)91004-X}

\bibitem{Luscher:2010iy}
M.~L{\"u}scher, JHEP \textbf{08}, 071 (2010).
\newblock \doi{10.1007/JHEP08(2010)071, 10.1007/JHEP03(2014)092}.
\newblock [Erratum: JHEP03,092(2014)]

\bibitem{Belavin:1975fg}
A.A. Belavin, A.M. Polyakov, A.S. Schwartz, {\relax Yu}.S. Tyupkin, Phys. Lett.
  \textbf{B59}, 85 (1975).
\newblock \doi{10.1016/0370-2693(75)90163-X}

\bibitem{Schafer:1996wv}
T.~Sch{\"a}fer, E.V. Shuryak, Rev. Mod. Phys. \textbf{70}, 323 (1998).
\newblock \doi{10.1103/RevModPhys.70.323}

\bibitem{DelDebbio:2004ns}
L.~Del~Debbio, L.~Giusti, C.~Pica, Phys. Rev. Lett. \textbf{94}, 032003 (2005).
\newblock \doi{10.1103/PhysRevLett.94.032003}

\bibitem{Ce:2014sfa}
M.~Ce, C.~Consonni, G.P. Engel, L.~Giusti, PoS \textbf{LATTICE2014}, 353 (2014)

\bibitem{Poppitz:2012nz}
E.~Poppitz, T.~Sch{\"a}fer, M.~{\"U}nsal, JHEP \textbf{03}, 087 (2013).
\newblock \doi{10.1007/JHEP03(2013)087}

\bibitem{Banks:1979yr}
T.~Banks, A.~Casher, Nucl. Phys. \textbf{B169}, 103 (1980).
\newblock \doi{10.1016/0550-3213(80)90255-2}

\bibitem{Ginsparg:1981bj}
P.H. Ginsparg, K.G. Wilson, Phys. Rev. \textbf{D25}, 2649 (1982).
\newblock \doi{10.1103/PhysRevD.25.2649}

\bibitem{Cherman:2016hcd}
A.~Cherman, T.~Sch{\"a}fer, M.~Unsal, {Chiral Lagrangian from Duality and
  Monopole Operators in Compactified QCD} (2016).
\newblock ArXiv:1604.06108

\bibitem{Beane:2003da}
S.R. Beane, P.F. Bedaque, A.~Parreno, M.J. Savage, Phys. Lett. \textbf{B585},
  106 (2004).
\newblock \doi{10.1016/j.physletb.2004.02.007}

\bibitem{Cristoforetti:2012su}
M.~Cristoforetti, F.~Di~Renzo, L.~Scorzato, Phys. Rev. \textbf{D86}, 074506
  (2012).
\newblock \doi{10.1103/PhysRevD.86.074506}

\bibitem{Aarts:2014nxa}
G.~Aarts, L.~Bongiovanni, E.~Seiler, D.~Sexty, JHEP \textbf{10}, 159 (2014).
\newblock \doi{10.1007/JHEP10(2014)159}

\bibitem{Aarts:2009uq}
G.~Aarts, E.~Seiler, I.O. Stamatescu, Phys. Rev. \textbf{D81}, 054508 (2010).
\newblock \doi{10.1103/PhysRevD.81.054508}

\bibitem{Sexty:2013ica}
D.~Sexty, Phys. Lett. \textbf{B729}, 108 (2014).
\newblock \doi{10.1016/j.physletb.2014.01.019}

\bibitem{Kloiber:2013rba}
T.~Kloiber, C.~Gattringer, PoS \textbf{LATTICE2013}, 206 (2014)

\bibitem{Schafer:2009dj}
T.~Sch{\"a}fer, D.~Teaney, Rept. Prog. Phys. \textbf{72}, 126001 (2009).
\newblock \doi{10.1088/0034-4885/72/12/126001}

\bibitem{Schaefer:2014awa}
T.~Sch{\"a}fer, Ann. Rev. Nucl. Part. Sci. \textbf{64}, 125 (2014).
\newblock \doi{10.1146/annurev-nucl-102313-025439}

\bibitem{Karsch:1986cq}
F.~Karsch, H.W. Wyld, Phys. Rev. \textbf{D35}, 2518 (1987).
\newblock \doi{10.1103/PhysRevD.35.2518}

\bibitem{Meyer:2007ic}
H.B. Meyer, Phys. Rev. \textbf{D76}, 101701 (2007).
\newblock \doi{10.1103/PhysRevD.76.101701}

\bibitem{Meyer:2007dy}
H.B. Meyer, Phys. Rev. Lett. \textbf{100}, 162001 (2008).
\newblock \doi{10.1103/PhysRevLett.100.162001}

\bibitem{Sakai:2007cm}
S.~Sakai, A.~Nakamura, PoS \textbf{LAT2007}, 221 (2007).
\newblock \doi{10.1063/1.2729742}.
\newblock [AIP Conf. Proc.893,5(2007)]

\bibitem{Aarts:2007wj}
G.~Aarts, C.~Allton, J.~Foley, S.~Hands, S.~Kim, Phys. Rev. Lett. \textbf{99},
  022002 (2007).
\newblock \doi{10.1103/PhysRevLett.99.022002}

\bibitem{Aarts:2007va}
G.~Aarts, PoS \textbf{LAT2007}, 001 (2007)

\bibitem{Aarts:2006wt}
G.~Aarts, C.~Allton, J.~Foley, S.~Hands, S.~Kim, Nucl. Phys. \textbf{A785}, 202
  (2007).
\newblock \doi{10.1016/j.nuclphysa.2006.11.148}

\bibitem{Meyer:2008gt}
H.B. Meyer, JHEP \textbf{08}, 031 (2008).
\newblock \doi{10.1088/1126-6708/2008/08/031}

\bibitem{Romatschke:2009ng}
P.~Romatschke, D.T. Son, Phys. Rev. \textbf{D80}, 065021 (2009).
\newblock \doi{10.1103/PhysRevD.80.065021}

\bibitem{Arnold:2000dr}
P.B. Arnold, G.D. Moore, L.G. Yaffe, JHEP \textbf{11}, 001 (2000).
\newblock \doi{10.1088/1126-6708/2000/11/001}

\bibitem{Kovtun:2004de}
P.~Kovtun, D.T. Son, A.O. Starinets, Phys. Rev. Lett. \textbf{94}, 111601
  (2005).
\newblock \doi{10.1103/PhysRevLett.94.111601}

\bibitem{Romatschke:2009im}
P.~Romatschke, Int. J. Mod. Phys. \textbf{E19}, 1 (2010).
\newblock \doi{10.1142/S0218301310014613}

\bibitem{Rezzolla:2013}
L.~Rezzolla, O.~Zanotti, \emph{Relativistic Hydrodynamics} (Oxford University
  Press, 2013)

\bibitem{Jeon:2015dfa}
S.~Jeon, U.~Heinz, Int. J. Mod. Phys. \textbf{E24}(10), 1530010 (2015).
\newblock \doi{10.1142/S0218301315300106}

\bibitem{Colella:1984}
P.~Colella, P.R. Woodward, J.\ Comp.\ Phys. \textbf{54}, 174 (1984)

\bibitem{Blondin:1993}
J.M. Blondin, E.A. Lufkin, Astrophys.\ J.\ Supp.\ Ser. \textbf{88}, 589 (1993)

\bibitem{Schafer:2010dv}
T.~Sch{\"a}fer, Phys. Rev. \textbf{A82}, 063629 (2010).
\newblock \doi{10.1103/PhysRevA.82.063629}

\bibitem{Florkowski:2010cf}
W.~Florkowski, R.~Ryblewski, Phys. Rev. \textbf{C83}, 034907 (2011).
\newblock \doi{10.1103/PhysRevC.83.034907}

\bibitem{Martinez:2010sc}
M.~Martinez, M.~Strickland, Nucl. Phys. \textbf{A848}, 183 (2010).
\newblock \doi{10.1016/j.nuclphysa.2010.08.011}

\bibitem{Bluhm:2015raa}
M.~Bluhm, T.~Sch{\"a}fer, Phys. Rev. \textbf{A92}(4), 043602 (2015).
\newblock \doi{10.1103/PhysRevA.92.043602}

\bibitem{Bluhm:2015bzi}
M.~Bluhm, T.~Sch{\"a}fer, Phys. Rev. Lett. \textbf{116}(11), 115301 (2016).
\newblock \doi{10.1103/PhysRevLett.116.115301}

\bibitem{Romatschke:2011hm}
P.~Romatschke, M.~Mendoza, S.~Succi, Phys. Rev. \textbf{C84}, 034903 (2011).
\newblock \doi{10.1103/PhysRevC.84.034903}

\bibitem{Brewer:2015hua}
J.~Brewer, M.~Mendoza, R.E. Young, P.~Romatschke, Phys. Rev. \textbf{A93}(1),
  013618 (2016).
\newblock \doi{10.1103/PhysRevA.93.013618}

\bibitem{Kadanoff}
L.P. Kadanoff, G.~Baym, \emph{Quantum Statistical Mechanics} (W.~A.~Benjamin,
  1962)

\bibitem{Bertsch:1988ik}
G.F. Bertsch, S.~Das~Gupta, Phys. Rept. \textbf{160}, 189 (1988).
\newblock \doi{10.1016/0370-1573(88)90170-6}

\bibitem{Lepers:2010be}
T.~Lepers, D.~Davesne, S.~Chiacchiera, M.~Urban, Phys. Rev. \textbf{A82},
  023609 (2010).
\newblock \doi{10.1103/PhysRevA.82.023609}

\bibitem{Bass:1998ca}
S.A. Bass, et~al., Prog. Part. Nucl. Phys. \textbf{41}, 255 (1998).
\newblock \doi{10.1016/S0146-6410(98)00058-1}.
\newblock [Prog. Part. Nucl. Phys.41,225(1998)]

\bibitem{Buss:2011mx}
O.~Buss, T.~Gaitanos, K.~Gallmeister, H.~van Hees, M.~Kaskulov, O.~Lalakulich,
  A.B. Larionov, T.~Leitner, J.~Weil, U.~Mosel, Phys. Rept. \textbf{512}, 1
  (2012).
\newblock \doi{10.1016/j.physrep.2011.12.001}

\bibitem{Ehehalt:1996uq}
W.~Ehehalt, W.~Cassing, Nucl. Phys. \textbf{A602}, 449 (1996).
\newblock \doi{10.1016/0375-9474(96)00097-8}

\bibitem{Geiger:1991nj}
K.~Geiger, B.~Muller, Nucl. Phys. \textbf{B369}, 600 (1992).
\newblock \doi{10.1016/0550-3213(92)90280-O}

\bibitem{Xu:2004mz}
Z.~Xu, C.~Greiner, Phys. Rev. \textbf{C71}, 064901 (2005).
\newblock \doi{10.1103/PhysRevC.71.064901}

\bibitem{McLerran:1993ni}
L.D. McLerran, R.~Venugopalan, Phys. Rev. \textbf{D49}, 2233 (1994).
\newblock \doi{10.1103/PhysRevD.49.2233}

\bibitem{Mueller:2002gd}
A.H. Mueller, D.T. Son, Phys. Lett. \textbf{B582}, 279 (2004).
\newblock \doi{10.1016/j.physletb.2003.12.047}

\bibitem{Mrowczynski:2016etf}
S.~Mrowczynski, B.~Schenke, M.~Strickland,   (2016)

\bibitem{Berges:2008wm}
J.~Berges, A.~Rothkopf, J.~Schmidt, Phys. Rev. Lett. \textbf{101}, 041603
  (2008).
\newblock \doi{10.1103/PhysRevLett.101.041603}

\bibitem{Dusling:2010rm}
K.~Dusling, T.~Epelbaum, F.~Gelis, R.~Venugopalan, Nucl. Phys. \textbf{A850},
  69 (2011).
\newblock \doi{10.1016/j.nuclphysa.2010.11.009}

\bibitem{Litim:2001db}
D.F. Litim, C.~Manuel, Phys. Rept. \textbf{364}, 451 (2002).
\newblock \doi{10.1016/S0370-1573(02)00015-7}

\bibitem{Wong:1970fu}
S.K. Wong, Nuovo Cim. \textbf{A65}, 689 (1970).
\newblock \doi{10.1007/BF02892134}

\bibitem{Hu:1996sf}
C.R. Hu, B.~Muller, Phys. Lett. \textbf{B409}, 377 (1997).
\newblock \doi{10.1016/S0370-2693(97)00851-4}

\bibitem{Maldacena:1997re}
J.M. Maldacena, Int. J. Theor. Phys. \textbf{38}, 1113 (1999).
\newblock \doi{10.1023/A:1026654312961}.
\newblock [Adv. Theor. Math. Phys.2,231(1998)]

\bibitem{Son:2007vk}
D.T. Son, A.O. Starinets, Ann. Rev. Nucl. Part. Sci. \textbf{57}, 95 (2007).
\newblock \doi{10.1146/annurev.nucl.57.090506.123120}

\bibitem{Gubser:2009md}
S.S. Gubser, A.~Karch, Ann. Rev. Nucl. Part. Sci. \textbf{59}, 145 (2009).
\newblock \doi{10.1146/annurev.nucl.010909.083602}

\bibitem{CasalderreySolana:2011us}
J.~Casalderrey-Solana, H.~Liu, D.~Mateos, K.~Rajagopal, U.A. Wiedemann,
  (2011)

\bibitem{DeWolfe:2013cua}
O.~DeWolfe, S.S. Gubser, C.~Rosen, D.~Teaney, Prog. Part. Nucl. Phys.
  \textbf{75}, 86 (2014).
\newblock \doi{10.1016/j.ppnp.2013.11.001}

\bibitem{Policastro:2002se}
G.~Policastro, D.T. Son, A.O. Starinets, JHEP \textbf{09}, 043 (2002).
\newblock \doi{10.1088/1126-6708/2002/09/043}

\bibitem{Teaney:2006nc}
D.~Teaney, Phys. Rev. \textbf{D74}, 045025 (2006).
\newblock \doi{10.1103/PhysRevD.74.045025}

\bibitem{Son:2006em}
D.T. Son, A.O. Starinets, JHEP \textbf{03}, 052 (2006).
\newblock \doi{10.1088/1126-6708/2006/03/052}

\bibitem{Chesler:2010bi}
P.M. Chesler, L.G. Yaffe, Phys. Rev. Lett. \textbf{106}, 021601 (2011).
\newblock \doi{10.1103/PhysRevLett.106.021601}

\bibitem{Chesler:2013lia}
P.M. Chesler, L.G. Yaffe, JHEP \textbf{07}, 086 (2014).
\newblock \doi{10.1007/JHEP07(2014)086}

\bibitem{Rangamani:2009xk}
M.~Rangamani, Class. Quant. Grav. \textbf{26}, 224003 (2009).
\newblock \doi{10.1088/0264-9381/26/22/224003}

\bibitem{Clay-YM}
A.~Jaffe, E.~Witten, {Quantum Yang Mills Theory, Official Description of
  Millenium Prize Problems} (2000).
\newblock {www.claymath.org}

\bibitem{Clay-NS}
C.~Fefferman, {Existence and smoothness of the Navier-Stokes Equation, Official
  Description of Millenium Prize Problems} (2000).
\newblock {www.claymath.org}

\bibitem{Villani:2009}
C.~Mouhot, C.~Villani, Acta Mathematica \textbf{207}, 29 (2011).
\newblock \doi{10.1007/s11511-011-0068-9}

\bibitem{Perelman}
G.~Perelman, {The entropy formula for the Ricci flow and its geometric
  applications} (2002).
\newblock ArXiv:math/0211159 [math.DG]

\bibitem{Shen:2014vra}
C.~Shen, Z.~Qiu, H.~Song, J.~Bernhard, S.~Bass, U.~Heinz, Comput. Phys. Commun.
  \textbf{199}, 61 (2016).
\newblock \doi{10.1016/j.cpc.2015.08.039}

\bibitem{Creutz:2004}
M.~Creutz, Computers in Science \& Engineering \textbf{March/April 2004}, 80
  (2004)

\end{thebibliography}
\end{document}